\documentclass[amsmath,amssymb, twocolumn,superscriptaddress,notitlepage,sort&compress,numbers]{revtex4-1}

\usepackage{comment}

\DeclareMathOperator\arctanh{arctanh}

\usepackage{upgreek}
\usepackage[dvipsnames]{xcolor}
\usepackage{url}
\usepackage{graphicx}
\usepackage{hyperref}
\hypersetup{colorlinks=true,
linkcolor = black,
urlcolor  = black,
citecolor = MidnightBlue,
pdfborder={0 0 0}}
\usepackage[utf8]{inputenc}
\usepackage[english]{babel}
\usepackage{amsfonts,amsmath,amssymb}
\usepackage{latexsym}

\usepackage{enumitem}

\newcommand{\latestChanges}[1]{{\color{Green} #1}}
\renewcommand{\latestChanges}[1]{#1}

\newcommand{\vecRho}{\boldsymbol \rho}

\usepackage{soul}
\setstcolor{Red}

\newcommand{\mean}[1]{\langle #1 \rangle}
\newcommand{\eq}[1]{
\begin{equation}
\begin{array}{c}
#1
\end{array}
\end{equation}
}
\renewcommand{\eqref}[1]{Eq. \ref{#1}}
\newcommand{\figref}[1]{Fig. \ref{#1}}
\newcommand{\Int}{
\displaystyle \int
}
\newcommand{\Sum}{
\displaystyle \sum 
}

\newcommand{\im}[1]{
\text{Im}\;#1
}
\newcommand{\re}[1]{
\text{Re}\;#1
}

\newcommand{\dbar}[1]{\Bar{\Bar{#1}}}
\newcommand{\G}{\dbar{\mathbf{G}}}

\usepackage{breqn}
\newcommand{\eqbreak}[1]{
\begin{dmath}
#1
\end{dmath}
}
\renewcommand{\vec}[1]{\mathbf{#1}}
\newcommand{\vhat}[1]{\hat{\mathbf{#1}}}
\newcommand{\mycomment}[1]{%
}%

\newcommand{\ket}[1]{\left|#1\right\rangle}

\begin{document}

\title{Maximum refractive index of an atomic medium}

\author{Francesco Andreoli}
\affiliation{\textit{ICFO - Institut de Ciències Fotòniques$,$ The Barcelona Institute of Science and Technology \\08860 Castelldefels$,$ Spain}}
\author{Michael J. Gullans}
\affiliation{Department of Physics$,$ Princeton University$,$ Princeton$,$ New Jersey 08544$,$ USA}
\author{Alexander A. High}
\affiliation{Pritzker School of Molecular Engineering$,$ University of Chicago$,$ Chicago$,$ Illinois 60637$,$ USA}
\affiliation{Center for Molecular Engineering and Materials Science Division$,$ Argonne National Laboratory$,$\\Lemont$,$ Illinois 60439$,$ USA}
\author{Antoine Browaeys}
\affiliation{Université Paris-Saclay$,$ Institut d’Optique Graduate School$,$ CNRS$,$ Laboratoire Charles Fabry$,$\\F-91127 Palaiseau$,$ France}
\author{Darrick E. Chang}
\affiliation{\textit{ICFO - Institut de Ciències Fotòniques$,$ The Barcelona Institute of Science and Technology \\08860 Castelldefels$,$ Spain}}
\affiliation{ICREA - Institució Catalana de Recerca i Estudis Avançats$,$ 08015 Barcelona$,$ Spain}

\begin{abstract}
It is interesting to observe that all optical materials with a positive refractive index have a value of index that is of order unity. Surprisingly, though, a deep understanding of the mechanisms that lead to this universal behavior seems to be lacking. Moreover, this observation is difficult to reconcile with the fact that a single, isolated atom is known to have a giant optical response, as characterized by a resonant scattering cross section that far exceeds its physical size. Here, we theoretically and numerically investigate the evolution of the optical properties of an ensemble of ideal atoms as a function of density, starting from the dilute gas limit, including the effects of multiple scattering and near-field interactions. Interestingly, despite the giant response of an isolated atom, we find that the maximum index does not indefinitely grow with increasing density, but rather reaches a limiting value $n\approx 1.7$. This limit arises purely from electrodynamics, as it occurs at densities far below that where chemical processes become important. We propose an explanation based upon strong-disorder renormalization group theory, in which the near-field interaction combined with random atomic positions results in an inhomogeneous broadening of atomic resonance frequencies. This mechanism ensures that regardless of the physical atomic density, light at any given frequency only interacts with at most a few near-resonant atoms per cubic wavelength, thus limiting the maximum index attainable. Our work is a promising first step to understand the limits of refractive index from a bottom-up, atomic physics perspective, and also introduces renormalization group as a powerful tool to understand the generally complex problem of multiple scattering of light overall.
\end{abstract}

\maketitle
\section{Introduction}
\begin{figure}[t!]
\centering
\includegraphics[width=\columnwidth]{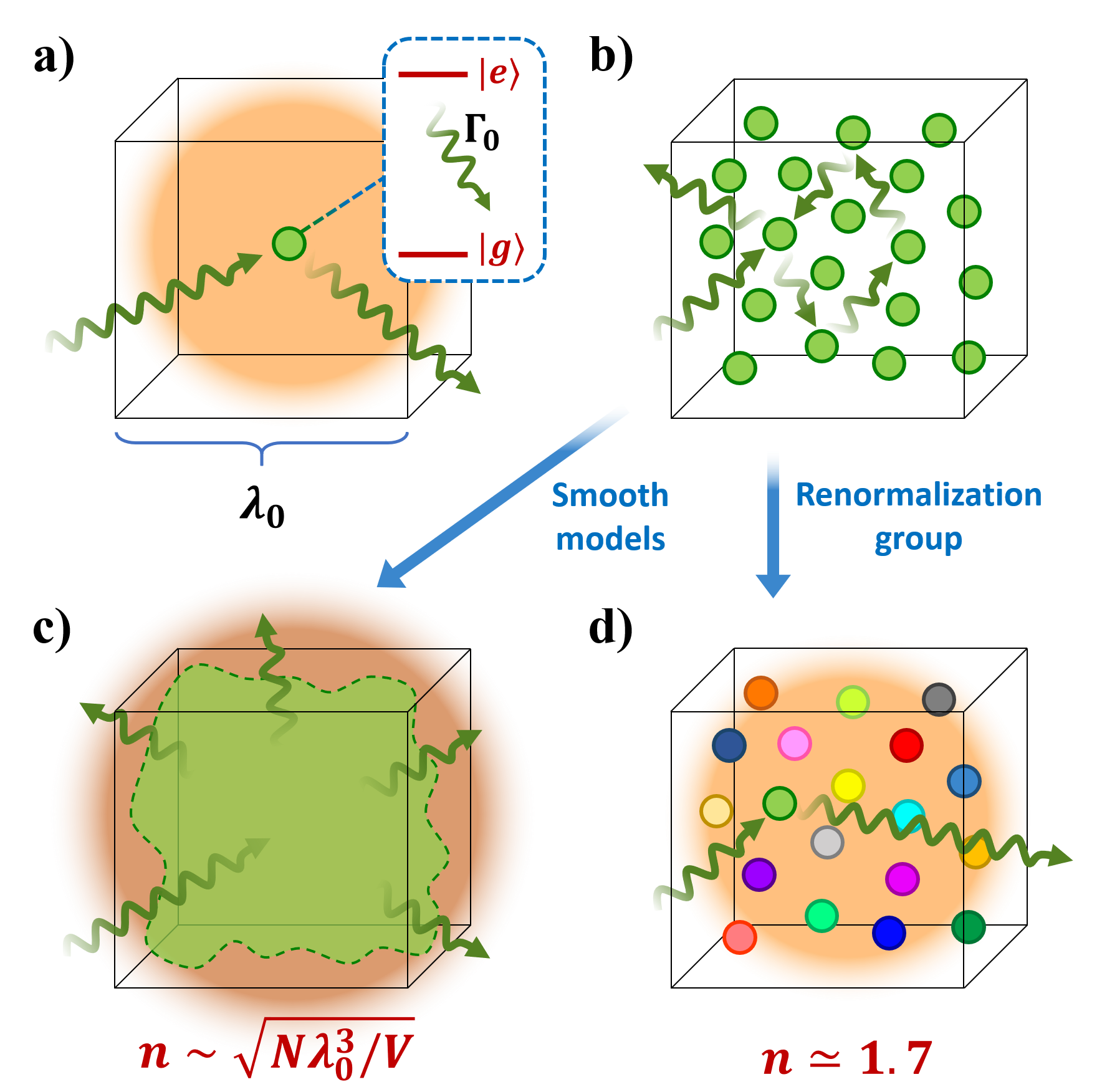}
\caption{\textbf{Optical response of an atomic medium.} a) Illustration of a single atom with a dipole-allowed optical transition between ground and excited states $\ket g$ and $\ket e$, characterized by a transition wavelength $\lambda_0$ and spontaneous emission rate $\Gamma_0$. Such an atom exhibits a scattering cross section (illustrated by the shaded region) of $\sigma_{\text{sc}}\sim \lambda_0^2$ for a single resonant photon (wavy green arrows). b) In a dense ensemble with many atoms per cubic wavelength $\lambda_0^3$, the scattering of an incident photon can involve multiple scattering and interference between atoms. c) In conventional theories of macroscopic optical response, the atoms are approximated by a smooth medium, and the index is derived from the product of single-atom polarizability and density. The maximum index $n$ near the atomic resonance then scales with atomic density like $n\sim \sqrt{N\lambda_0^3/V}$. d) In our renormalization group theory, we retain multiple scattering and granularity, showing that the optical properties of the ensemble are determined by a hierarchy of nearby atomic pairs that strongly interact via their near fields. These interactions effectively produce an inhomogeneously broadened ensemble, where the amount of broadening scales with density (with the different colors of atoms representing the different resonance frequencies in the figure). An incident photon of a given frequency thus sees only $\sim 1$ near-resonant atom per reduced cubic wavelength to interact with, regardless of atomic density. This results in a maximum index of $n\approx 1.7$.}
\label{fig:scatt_cross_sec}
\end{figure}

One interesting observation is that all the optical materials that we know of, with a positive index of refraction at visible wavelengths, universally have an index of order unity, $n\sim \mathcal O(1)$. While we typically utilize materials far from their natural electronic resonances, this even holds true close to resonance \cite{Rustgi1961OpticalEv,Philipp1963OpticalSemiconductors,Walker1964UltravioletDiamond,Lamy1977OpticalUltraviolet,Aspnes1983DielectricEV,Warren1984OpticalMicrowave,Papadopoulos1991OpticalDiamond,Kitamura2007OpticalTemperature}. Yet, despite the profound implications that an ultra-high index material would have for optical technologies, a deep understanding of the origin of this apparently universal behavior seems to be lacking. Furthermore, this property of real materials is not readily reconciled with the fact that a single, isolated atom exhibits a giant scattering cross-section $\sigma_{\text{sc}} \sim\lambda_0^2$ for photons resonant with an atomic transition of wavelength $\lambda_0$ (\figref{fig:scatt_cross_sec}-a), which far exceeds both the physical size of the atom or the typical lattice constant of a solid ($\lambda_0\sim 1\upmu \text{m}$ for a typical optical transition, compared to the Bohr radius $a_0\sim 0.1 \text{nm}$).

In standard theories \cite{Jackson1998ClassicalElectrodynamics,Grynberg2010IntroductionOptics}, the macroscopic index of an atomic medium (\figref{fig:scatt_cross_sec}-b) is constructed from the product of the single-atom polarizability and the atomic density, and around resonance its value $n\sim \sqrt{N\lambda_0^3/V}$ extrapolates to a maximum of $ \sim 10^5$ at solid densities (\figref{fig:scatt_cross_sec}-c). It is well-known that this argument neglects multiple scattering of light and photon-mediated dipole-dipole interactions \cite{Fleischhauer1999RadiativeFormula,Javanainen2016LightOptics}, and substantial work has been devoted to explore their effects on various optical phenomena, such as 
collective resonance shifts 
\cite{ Manassah2012CooperativeShift,Keaveney2012CooperativeThickness,Javanainen2014ShiftsSample, Bromley2016CollectiveMedium, Jenkins2016OpticalGases,Dobbertin2020CollectiveNanocavities}, 
cooperative scattering properties
\cite{Schilder2016PolaritonicAtoms,Schilder2017HomogenizationScatterers,Schilder2020Near-ResonantAtoms}, 
emergence of sub- and super-radiance
\cite{Dicke1954CoherenceProcesses,Gross1982Superradiance:Emission, Roof2016ObservationAtoms,Araujo2016SuperradianceRegime,Asenjo-Garcia2017ExponentialArrays,He2019GeometricEmission}, 
%
realization of atomic mirrors 
\cite{Shahmoon2017CooperativeArrays,Bettles2016EnhancedArray, Rui2020ALayerb},
and Anderson localization of light 
\cite{Skipetrov2014AbsenceScatterers,Skipetrov2019SearchField}. 
In particular, this includes theoretical and experimental evidence that the optical response of dense gases can be much smaller than standard predictions %
\cite{Pellegrino2014ObservationEnsemble,Schilder2016PolaritonicAtoms, Jennewein2018CoherentTheory,Javanainen2016LightOptics, Guerin2017LightEffects} 
or even reach limiting values \cite{Chomaz2012AbsorptionAnalysis, Zhu2016LightMedia,Jenkins2016CollectiveExperiment,Jennewein2016PropagationAtoms,Corman2017TransmissionAtoms,Jennewein2016CoherentCloud,Keaveney2012MaximalNanolayer}.
However, an underlying physical explanation is still missing, and our goal here is to understand better the mechanisms that might limit the index even when operating close to resonance.

Specifically, we investigate in detail the optical response of an ideal ensemble of identical, stationary atoms, as a function of density starting from the dilute limit, and well within the regime where the atoms do not interact chemically. In large scale numerics (involving up to $\sim 23000$ atoms, about an order of magnitude larger than comparable works \cite{Chomaz2012AbsorptionAnalysis,Javanainen2014ShiftsSample,Javanainen2016LightOptics,Zhu2016LightMedia, Schilder2016PolaritonicAtoms,Schilder2017HomogenizationScatterers,Schilder2020Near-ResonantAtoms, Jennewein2016PropagationAtoms,Jennewein2018CoherentTheory,Corman2017TransmissionAtoms,Jenkins2016CollectiveExperiment,Dobbertin2020CollectiveNanocavities}), we find that the maximum index does not indefinitely grow with density, and saturates to a maximum value of $n\approx 1.7$, when the typical distance between atoms becomes smaller than the length scale associated with the resonant cross section, i.e. $d<\lambda_0$. Furthermore, we introduce an underlying theory based upon strong-disorder renormalization group (RG), which has been a very successful technique to deal with highly varying interaction strengths in a wide variety of condensed matter systems \cite{Levitov1990DelocalizationInteraction,Fisher1994RandomChains,Damle2000DynamicsChains,Motrunich2000Infinite-randomnessPoints,Refael2004EntanglementDimension,Igloi2005StrongSystems,Vosk2013Many-bodyPoint,Refael2013StrongTransition}. In the context of our particular problem, the combination of strong near-field ($\sim 1/r^3$) optical interactions and random atomic positions enables one to characterize the optical response of the system in terms of a hierarchy of strongly interacting, nearby atomic pairs. The shifts of the resonance frequencies arising from the near-field interactions then effectively yield an inhomogeneously broadened optical medium, where the amount of broadening linearly scales with density. This implies that light of any given wavelength only interacts with at most $\sim 1$ near-resonant atom per reduced cubic wavelength $\lambda_0^3/(2\pi)^3$, regardless of the physical atomic density, thus limiting the optical response (\figref{fig:scatt_cross_sec}-d).

Our results are potentially significant on a number of fronts. First, they provide a convincing picture of why typical theories for optical response, based upon a smooth density approximation, fail for dense, near-resonant atomic media, due to the important role of granularity and strong interactions of any given atom with a particularly close-by, single neighbour. Furthermore, our results show the promise of a bottom-up approach to understanding the physical limits of refractive index, starting from objects (isolated atoms) whose optical responses are both huge and exquisitely understood. Separately, the existence of a fundamental mechanism that results in inhomogeneous broadening (i.e. dephasing) and saturation of optical properties at high densities, which occurs even for perfect, stationary atoms, should impose fundamental bounds on the maximum densities and minimum sizes of atom-light interfaces needed to realize high-fidelity quantum technologies. Finally, while we focus here on the linear optical response of a dense atomic medium, we believe that the validity of RG is quite general, and can constitute a versatile new tool for the generally challenging problem of multiple scattering in near-resonant disordered media 
\cite{Chomaz2012AbsorptionAnalysis,Lagendijk1996ResonantLight,Skipetrov2014AbsenceScatterers,Schilder2017HomogenizationScatterers,Javanainen2016LightOptics,Fayard2015IntensityPatterns,Guerin2017LightEffects,Corman2017TransmissionAtoms,Zhu2016LightMedia,Jennewein2018CoherentTheory,Cottier2019MicroscopicLight}, 
including in the nonlinear and quantum regimes \cite{Binninger2019NonlinearCloud}.

This paper is structured as follows. First, we briefly review the theoretical formulation of the multiple scattering problem of atoms or other point-like dipoles, and the standard atomic physics model of refractive index, when atomic granularity and multiple scattering are ignored. We then formulate our large-scale numerical simulations, describing a few implementation details that allow the index to be efficiently calculated, and show that the index eventually saturates with increasing density to a maximum value of $n\approx 1.7$. We then introduce our RG theory, which highlights the importance of granularity and nearby atomic pairs on the macroscopic optical response, before concluding with an expanded discussion of future interesting directions to investigate.
\section{Formal theory of multiple scattering}
We consider a minimal system consisting of $N$ identical, stationary two-level atoms. The atoms are assumed to have an electronic ground and excited state $\ket g$, $\ket e$, with frequency difference $\omega_0$ and associated wavelength $\lambda_0=2\pi c/\omega_0$, and which have an electric dipole transition with a dipole matrix element along a fixed axis (say $\vhat{x}$), as depicted in \figref{fig:scatt_cross_sec}-a. The excited states of the atoms decay purely radiatively, with a rate of $\Gamma_0$ for a single, isolated atom. As we are specifically interested in the linear refractive index, it is sufficient to treat atoms in the limit of classical, polarizable, radiating dipoles. In order to investigate the frequency-dependent index $n(\omega)$, we consider that the atoms are driven by a monochromatic, linearly-polarized input beam $\vec E_{\text{in}}(\vec r,\omega)=E_{\text{in}}(\vec r,\omega)\vhat x$, whose polarization aligns with the polarizability axis of the atoms. Each atom $j$ acquires a dipole moment $\vec d_j(\omega)=d_j(\omega)\vhat x$, as a result of being driven by the total field, which consists of the sum of the incident field and fields re-scattered from other atoms. Formally, the total field can be expressed as \cite{Novotny2009PrinciplesNano-optics}
\eq{
\label{eq:FTMS_output_field}
\vec E(\vec r, \omega)= \vec E_{\text{in}}(\vec r, \omega) +\mu_0 \omega^2\Sum_{j=1}^N \G(\vec r, \vec r_j, \omega)\cdot \vec d_j(\omega).
}

\begin{figure}[t!]
\centering
\includegraphics[width=1\columnwidth]{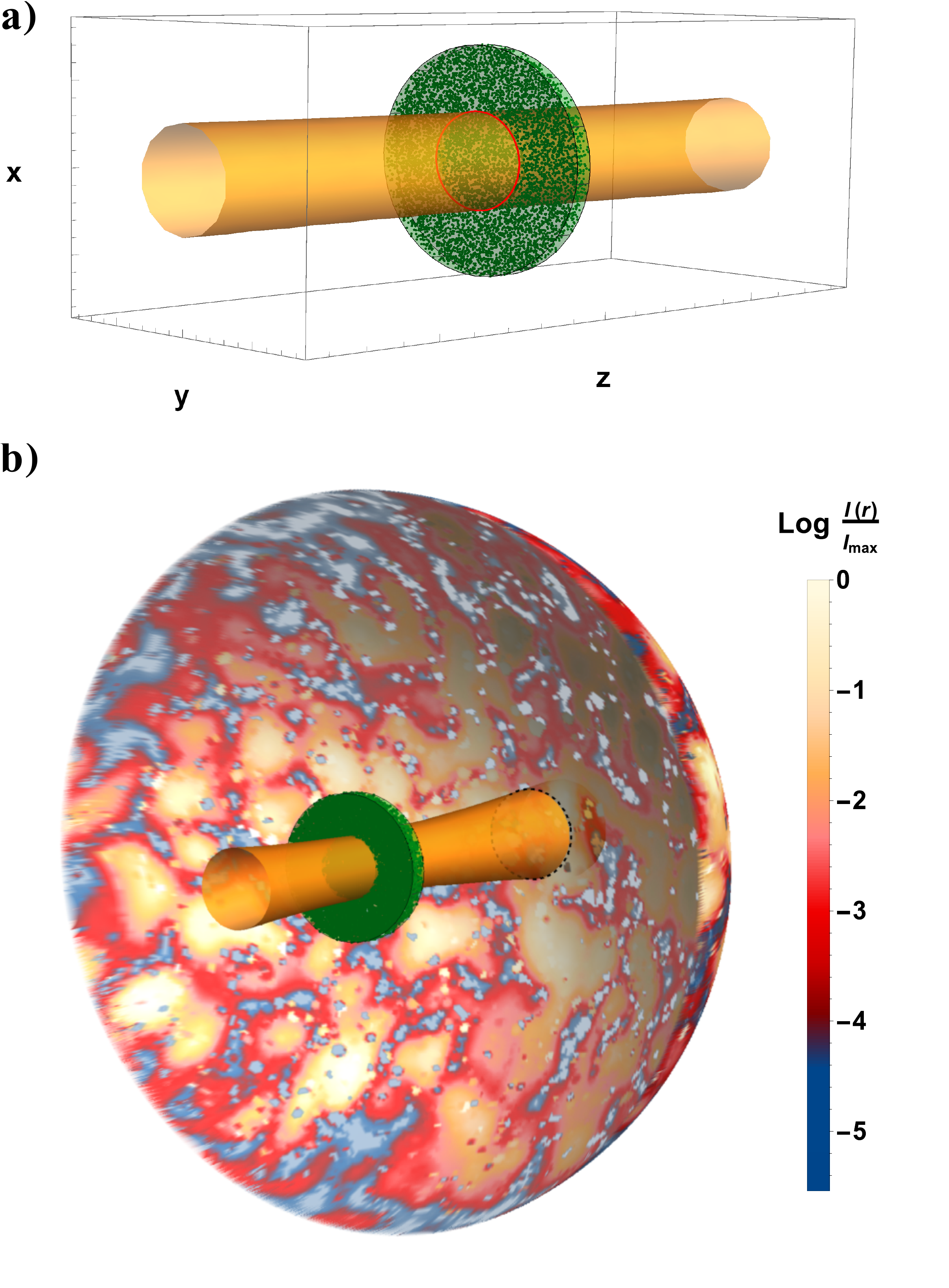}
\caption{\textbf{Simulated physical system.} a) A cylindrical ensemble of randomly distributed atoms (green points) is illuminated by a \textit z-directed Gaussian beam, whose beam waist $w(z)\gg \lambda_0$ is represented in orange. The transverse radius of the cylinder is chosen to be much larger than the beam waist, to avoid edge diffraction.
b) Color-coded 3D representation of the forward scattered intensity \latestChanges{$I(\vec r,\omega_0)=|\vec E(\vec r,\omega_0)|^2/(2\mu_0 c)$} (with the value indicated in the colorbar) over a hemispherical surface far from the ensemble (the radius of this hemisphere is $35\lambda_0$), given an incident resonant Gaussian beam. The intensity is calculated for a single, random atomic configuration. The system parameters used are: beam waist $w_0=3\lambda_0$, cylinder radius $l_{\text{cyl.}}=7 \lambda_0$ and thickness $d=2\lambda_0$.
}
\label{fig:system_representation}
\end{figure}

Here, the dyadic Green's tensor $\G(\vec r, \vec r_j, \omega)$ encodes the field at position $\vec r$, produced by an oscillating dipole at $\vec r_j$, and in vacuum is given by \cite{Novotny2009PrinciplesNano-optics}
\eqbreak{
\label{eq:FTMS_Green_tensor}
\G (\vec r,\vec r',\omega) =
k \dfrac{e^{i \rho}}{4\pi }\left[\left(\dfrac{1}{\rho}+\dfrac{i}{\rho^2}-\dfrac{1}{\rho^3}\right)\mathbb I\\\\
+\left(-\dfrac{1}{\rho}-\dfrac{3i}{\rho^2}+\dfrac{3}{\rho^3}\right)\dfrac{\boldsymbol{\rho} \otimes \boldsymbol{\rho}  }{\rho^2}\right],
}
with dimensionless distance defined as $\rho \equiv |\boldsymbol\rho|\equiv k| (\vec r - \vec r')|$ and $k=\omega/c$. Note that $\G (\vec r,\vec r',\omega)$ contains both non-radiative, near-field ($\sim 1/\rho^3$) and radiative, far-field ($\sim 1/\rho$) terms. 
Then, the induced dipole moment of atom $i$ is given by
\eqbreak{
\label{eq:FTMS_coupled_dipoles}
 d_i(\omega)=\alpha_0(\omega) \epsilon_0\left[ E_{\text{in}}(\vec r_i,\omega)
\\\\
+\mu_0 \omega^2 \Sum_{j\neq i}^{N-1} \vhat x \cdot \G(\vec r_i,\vec r_j, \omega)\cdot \vhat x\;  d_j(\omega)\right],}
where the parameter $\alpha_0(\omega)$ defines the polarizability of a single dipole. Although \eqref{eq:FTMS_output_field} and \eqref{eq:FTMS_coupled_dipoles} can describe any system of linearly-polarizable point-like dipoles \cite{Novotny2009PrinciplesNano-optics}, e.g. dielectric nano-particles \cite{GarciaDeAbajo2007CollectiveMatter}, in our case we focus on the response of non-absorbing, purely radiative atoms, whose resonant cross section $\sigma_{\text{sc}}=3\lambda_0^2/(2\pi)$ is the maximum set by the unitarity limit \cite{DeVries1998PointWaves}. In this context, the atomic polarizability reads $\alpha_0(\omega)=-3\pi /[(\Delta + i/2)k_0^3]$, where $k_0=2\pi/\lambda_0$ denotes the resonant wavevector, while $\Delta\equiv(\omega - \omega_0)/\Gamma_0$ represents the dimensionless detuning between the input beam frequency $\omega$ and the atomic resonance $\omega_0$. \latestChanges{To relate to other work, we note that an identical equation to \eqref{eq:FTMS_coupled_dipoles} can also be derived starting from a quantum mechanical formulation of atom-light interactions in the presence of multiple scattering, where the light-mediated interactions between atoms are encoded in a non-Hermitian Hamiltonian describing dipole-dipole interactions \cite{Gross1982Superradiance:Emission}. More precisely, one can focus on the regime where at most one atom is excited, which reflects the low-intensity limit of linear optics that we are interested in. Then, the steady-state wave function amplitudes for atom $i$ to be excited obey the same coupled equations of \eqref{eq:FTMS_coupled_dipoles} \cite{Chomaz2012AbsorptionAnalysis,Skipetrov2014AbsenceScatterers,Bromley2016CollectiveMedium,Zhu2016LightMedia,Asenjo-Garcia2017ExponentialArrays}.}

While \eqref{eq:FTMS_output_field} and \eqref{eq:FTMS_coupled_dipoles} are formally exact, solving a number of equations that explicitly scales with the number of atoms and that depends on the details of atomic positions is not a particularly convenient way to calculate the index or other optical properties. Historically, this fostered the development of simplified theories for the macroscopic response, such as the Drude-Lorentz model \cite{Jackson1998ClassicalElectrodynamics} or equivalently the Maxwell-Bloch (MB) equations \cite{Grynberg2010IntroductionOptics}, where the discreteness of atoms is replaced by a smooth medium of density $N/V$ (\figref{fig:scatt_cross_sec}-c). The resulting index depends on the product of density and single-atom polarizability,
\eq{
\label{eq:FTMS_index_prediction_MB}
n_{\text{MB}}(\Delta)=\sqrt{1+\dfrac{N}{V}\alpha_0(\omega)}= \sqrt{1+ \dfrac{3\pi\eta}{-\Delta - i/2}},
}
where we defined the dimensionless density $\eta \equiv N/(Vk_0^3)$. Notably, for an optimum detuning, the maximum real part of the index scales as $\sim\sqrt{\eta}$. 

While the MB equations ignore multiple scattering, the Lorentz-Lorenz (LL) or the equivalent Clausius-Mossotti model is one well-known approach to approximate its effects, still within the smooth density approximation. Given any atom located at $\vec r_0$, the model approximates the neighbouring atoms as a smooth dielectric medium with a small spherical exclusion around $\vec r_0$ \cite{Jackson1998ClassicalElectrodynamics}. The resulting local field correction produced by the other atoms gives an index that satisfies the equation $(n_{\text{LL}}^2-1)/(n_{\text{LL}}^2+2) = (N/V)\alpha_0(\omega)/3$ \cite{Novotny2009PrinciplesNano-optics}. Plugging in the atomic polarizability, one readily finds that
\eq{
\label{eq:FTMS_index_prediction_LL}
n_{\text{LL}}(\Delta)=n_{\text{MB}}(\Delta+\pi\eta).}
Importantly, while the spectrum is shifted with respect to the MB model, the LL model still produces a maximum index that grows like $\sim\sqrt{\eta}$.
\begin{figure*}[t!]
\centering
\includegraphics[width=2\columnwidth]{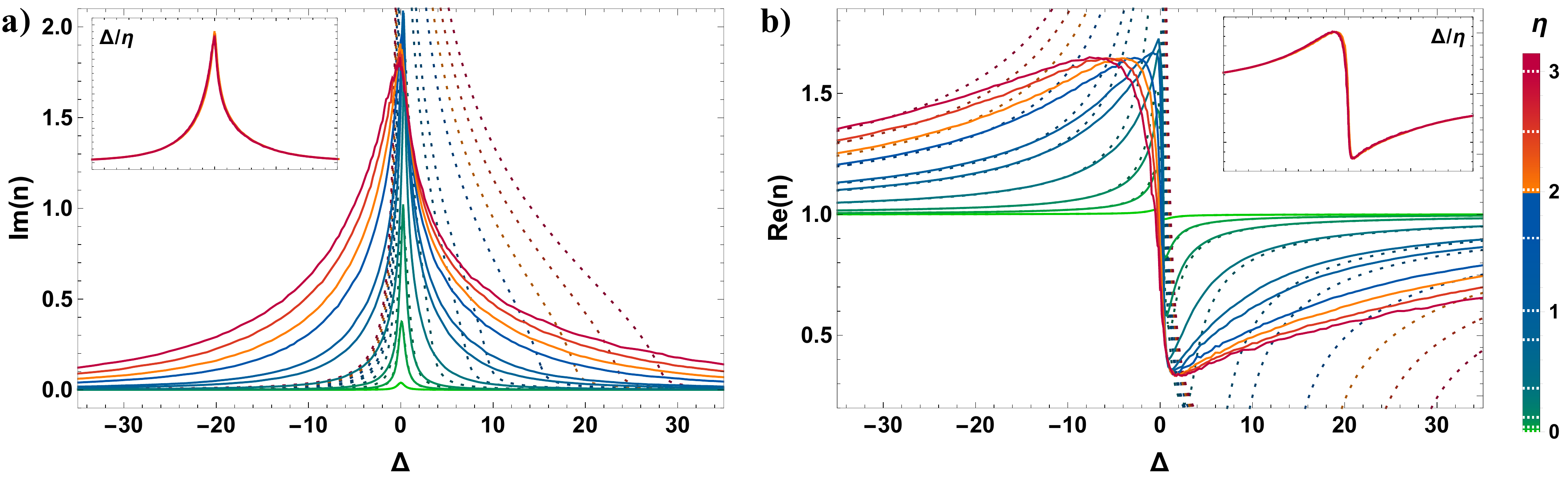}
\caption{
\textbf{Frequency-dependent refractive index for different atomic densities.} The solid lines portray the imaginary (subfigure $a$) and real (subfigure $b$) part of the refractive index versus dimensionless detuning $\Delta$, obtained through \eqref{eq:CDS_input-output_eq_single_mode}, while the dotted lines show the MB predictions. The colors denote different atomic densities (colorbar on right), with the specific values indicated by the dotted white lines. The refractive index is inferred by averaging the complex transmission coefficient $t(\Delta)$ over $\sim 10^3 - 10^4$ atomic configurations. Other system parameters are: thickness $d =0.4\lambda_0$, transverse radius $5 \leq l_{\text{cyl.}}/\lambda_0\leq 7$, beam waist $2.5\leq w_0/\lambda_0 \leq 3$. The insets show the curves at the 3 highest densities as a function of the rescaled detuning $\Delta/\eta$.
}
\label{fig:SM_spectrum}
\end{figure*}
\section{Coupled-dipole simulations}
\eqref{eq:FTMS_output_field} and \eqref{eq:FTMS_coupled_dipoles} are ubiquitously used to model multiple scattering and interference effects involving a moderate number of point-like scatterers. Here, we briefly introduce some key details of our implementation, which allows us to perform simulations on very high atom number and efficiently extract the index.

First, one conceptually straightforward way to extract the complex refractive index of a material would be to take a slab of thickness $d$ and large transverse extent, and investigate the phase shift and attenuation of a quasi-plane-wave incident field upon transmission. We approximately realize such a situation by taking atoms with a fixed density in a cylindrical \latestChanges{volume centered around the origin, illuminated by a weakly focused, near-resonant Gaussian beam. 
Decomposing the position $\vec{r}=\{\vec{r}_{\perp},z\}$ in terms of a transverse component $\vec{r}_{\perp}$ and axial component $z$, the beam amplitude within the paraxial approximation is given by $E_{\text{in}}(\vec r,\omega_0)=E_0 [w_0/w(z)]\exp\{-[r_{\perp}/w(z)]^2+i[k_0 z + \phi(\vec r,w_0)]\}$, where $w(z)=w_0\sqrt{1+(z/ z_{\text{R}})^2}$ describes the transverse extension of the beam, while $w_0=w(0)$ is the beam waist at the focal plane and $\phi(\vec r,w_0)$ accounts for the curvature of the wave-front and for the Gouy phase \cite{Grynberg2010IntroductionOptics}
(see \figref{fig:system_representation}-a for an illustration of the system). This parameter is given by $\phi(\vec r,w_0) = - \arctan(z/z_{\text{R}}) +  k_0 r_{\perp}^2 /\{2 z[1+(z_{\text{R}}/z)^2]\} $, with $z_{\text{R}}=k_0w_0^2/2$.
Given that the intensity of the beam drops off rapidly for transverse distances larger than $w(z)$, the parameters are chosen such that $w(z)$ is small compared to the radius of the cylinder, so that diffraction effects from the edges are negligible. Note that the cylindrical geometry has the nice feature that the furthest atoms are equidistant from the center of the beam. This avoids ``wasting'' computational resources, such as in a rectangular geometry, on atoms at the corners that hardly contribute to the optical response. Finally we avoid very tight focusing $w_0\lesssim \lambda_0$, where non-paraxial effects could emerge.} 

We must also specify a practical definition of index, for a granular system as ours. In particular, since our atoms are purely scattering and have no absorption, it is well-known \cite{Shapiro1986LargeMedia,Lagendijk1996ResonantLight,Fayard2015IntensityPatterns,Schilder2017HomogenizationScatterers} 
that for a fixed random spatial configuration, an input as in \figref{fig:system_representation} produces a complex ``speckle" pattern in the outgoing intensity when the system is optically dense, due to multiple scattering and interference, as exemplified in \figref{fig:system_representation}-b. To isolate the part of the field that possesses a well-defined phase relationship with the incident field from realization to realization, we project \eqref{eq:FTMS_output_field} back into the same Gaussian mode as the input, as can be experimentally enforced by recollecting the transmitted light through a single mode fiber. This results in a transmission coefficient $t(\Delta)$ given by \cite{Chomaz2012AbsorptionAnalysis,Manzoni2018OptimizationArrays}

\eq{
\label{eq:CDS_input-output_eq_single_mode}
t(\Delta)=1+\dfrac{3 i}{(w_0k_0)^2 }\Sum_{j=1}^N \dfrac{  E^*_{\text{in}}(\vec r_j,\omega_0)}{E_0}  c_j(\Delta),}
where $E_0$ is the input field amplitude at the beam focus. Here, for convenience, we have defined re-scaled dipole amplitudes $c_j(\Delta) = d_j(\omega) k_0^3/(3\pi \epsilon_0 E_0) $, which satisfy the dimensionless coupled equations
\eq{
\label{eq:CDS_input-output_eq_single_mode2}
-\Delta c_i(\Delta)-\displaystyle
\Sum_{j=1}^{N}  G_{ij} c_j(\Delta)   = \; \dfrac{ E_{\text{in}}(\vec r_i,\omega_0)}{E_0},
}
In these equations, we define $G_{ij}\equiv ( 3\pi/k_0) \vhat x \cdot \G(\vec r_i,\vec r_j,\omega_0)\cdot \vhat x$ and $G_{jj}=i/2$, which coincides with the single-atom decay rate in units of $\Gamma_0$, while regularizing the divergent self-energy associated with the real part of $\G$. Note that, for simplicity, the Green's function $\G(\vec r_i,\vec r_j,\omega_0)$ is only evaluated at the atomic resonance frequency, in order to ease the computational cost as the detuning is varied. Ignoring the dispersion of $\G$ is an excellent approximation for near-resonant atoms, as the optical dispersion and delay of such a system is dominated by the atomic response itself rather than from the vacuum \cite{Chang2012CavityMirrors}. Similarly, we approximate the near-resonant input field as $E_{\text{in}}(\vec r_i,\omega)\simeq E_{\text{in}}(\vec r_i,\omega_0)$.

The expression in \eqref{eq:CDS_input-output_eq_single_mode} represents a useful closed-form definition of the transmission coefficient $t(\Delta)$, which avoids a numerically expensive point-by-point evaluation of the scattered field $\vec E(\vec r,\omega)$, as nominally prescribed by \eqref{eq:FTMS_output_field}. We can extrapolate the complex index of refraction $n(\Delta)$ from the relation
\eq{
\label{eq:CDS_refractive_index_def_from_t}
\mean{t(\Delta)}= \exp \left\{ i  \left[n\left(\Delta \right)-1\right] k_0 d\right\}.
}
where the averages are performed over $\sim 10^3 - 10^4$ sets of random positions, for each fixed density. Unlike in a smooth medium, we have that $|\mean{ t(\Delta)}|^2 \neq \mean{|t(\Delta)|^2}$. Nevertheless, our definition of the index coincides with that often used within atomic physics (e.g. in phase contrast or absorption imaging of a Bose-Einstein condensate \cite{Davis1995Bose-EinsteinAtoms, Anderson1995ObservationVapor}). In Appendix A, we demonstrate the independence of the calculated index from the thickness $d$, which is implicitly assumed in \eqref{eq:CDS_refractive_index_def_from_t}. Alternatively, one might assume that the calculated $\mean{ t(\Delta)}$ approximately coincide with the finite-slab Fresnel coefficients for a smooth material \cite{Born1999PrinciplesOptics}.
This produces an alternative way to extrapolate the index, which we find yields quantitatively similar results as what we present below.

\begin{figure*}[t!]
\centering
\includegraphics[width=2\columnwidth]{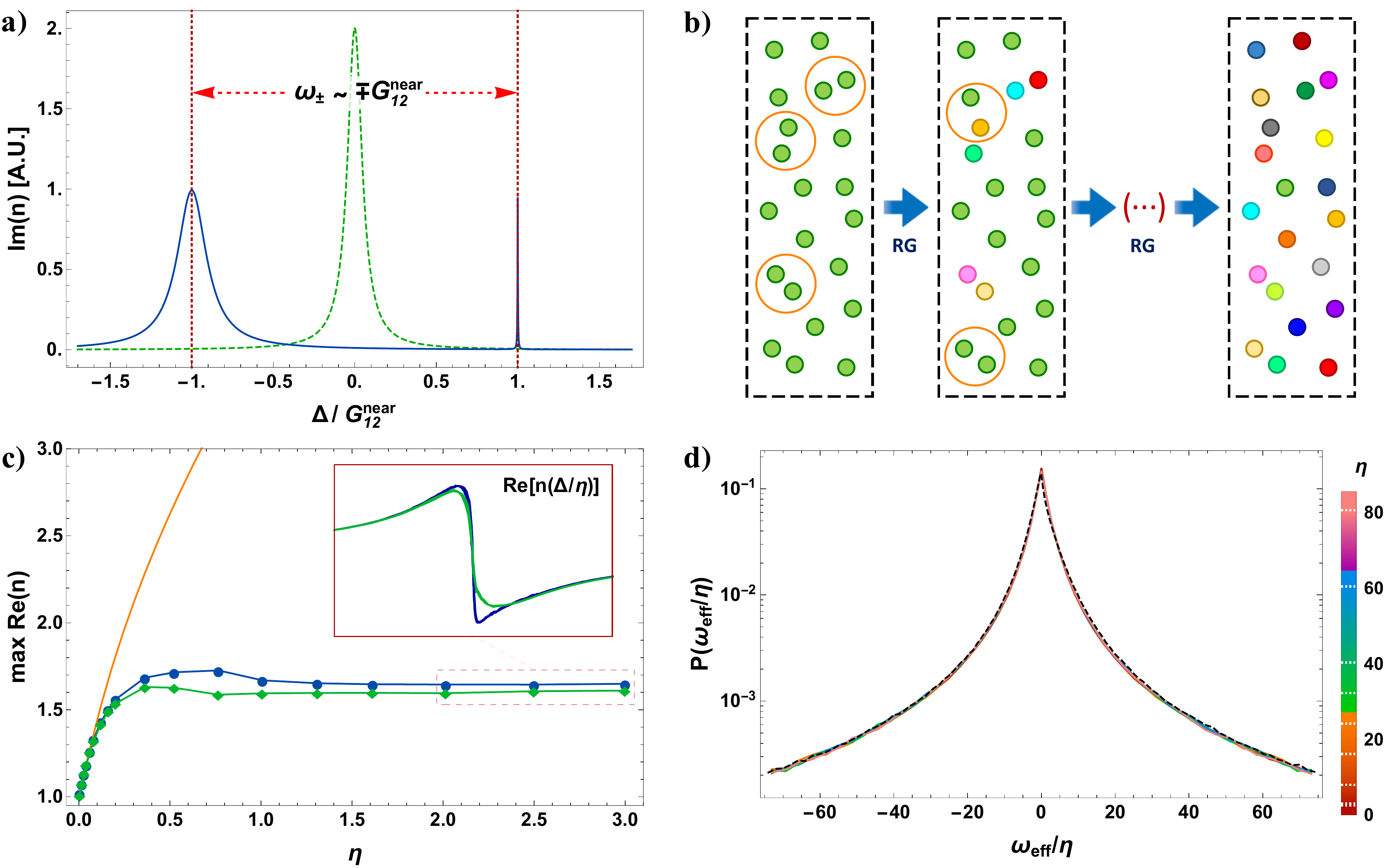}
\caption{\textbf{Renormalization group analysis.} a) Representative optical response of two identical atoms separated by a distance $\rho_{12} \ll 1$. Here, we plot the absorption spectrum (blue curve), which consists of two well-separated Lorentzians. The positions of the resonances are given by  $\sim \mp G_{12}^{\text{near}}$, where $G_{12}^{\text{near}}\propto 1/\rho_{12}^3$ is the near-field component of the Green's function. To compare, we also plot twice the response of a single, isolated atom (green dashed line). 
b) Pictorial representation of the RG scheme \latestChanges{for a many-atom system}. At each step of the RG flow the nearby pairs (identified by orange circles) that mostly strongly interact via their near fields are identified, and replaced with atoms with different resonance frequencies (indicated by different colors) in such a way to produce an equivalent optical response. At the end of the RG process (last panel) the overall system is equivalent to an inhomogeneously broadened ensemble of weakly interacting atoms.
c) Comparison between the maximum real refractive index predicted by the full coupled-dipole simulations of identical atoms (blue points), and index of the equivalent, inhomogeneously broadened ensemble predicted by RG (green). For each value of density, the maximum index is obtained by optimizing over detuning. For comparison, the MB and LL models both predict a maximum index given by the orange curve. The inset compares the rescaled spectra $\re n(\Delta/\eta)$ of the RG (green) and full coupled-dipole (blue) simulations, given the points at densities $\eta \gtrsim 2$.
d) Rescaled probability distribution of effective, inhomogeneously broadened resonance frequencies $P(\omega_{\text{eff}}/\eta)$ obtained from the application of the RG scheme. Given 9 different values of the density $\eta$ (ranging from $\eta \approx 2.5$ up to $\eta \approx 80$), the distributions of effective resonance frequencies are plotted with a different color, according to the bar on the right. The exact values chosen for the curves are emphasized by dotted white lines in the color-bar. \latestChanges{The curves at $\eta \approx 2.5$ and $\eta \approx 3$ are calculated using the cylindrical system studied in \figref{fig:SM_spectrum} (with thickness $d =0.4\lambda_0$ and transverse radius $l_{\text{cyl.}}= 5\lambda_0$), while the distributions $P(\omega_{\text{eff}}/\eta)$ at densities $\eta >3$ are evaluated using a spherical geometry of radius $r_{\text{sph.}}=0.55\lambda_0$. Finally, for the case of density $\eta\approx 3$ we plot (black dashed curve) the many-atom distribution of the eigenvalues of the near-field matrix $-G_{\text{near}}$ (also rescaled by a factor of $1/\eta$ for consistency), as discussed further in Sec. \ref{sec:M.J.RG.S.}. All distributions are obtained by accumulating results from $\sim 100$ different configurations of atomic positions.}
}
\label{fig:RG_plots}
\end{figure*}

In \figref{fig:SM_spectrum}, we plot our numerical results for the real and imaginary parts of $n(\Delta)$, as a function of the input field detuning $\Delta$, and for various densities. For comparison, we also plot the index as predicted by the MB equations, which starts to appreciably deviate from the full numerical results for dimensionless densities $\eta \gtrsim 0.1$. Interestingly, for sufficiently high densities, we observe that the computed spectra collapse onto the same curve when plotted as a function of the re-scaled detuning $\Delta/\eta$, as shown in the insets of \figref{fig:SM_spectrum}, which include all plots in the range $ 2 \lesssim \eta\lesssim 3$. The invariance of $n(\Delta/\eta)$ for $\eta \gtrsim 2$ directly indicates that both the maximum real index and the attenuation per unit length acquire fixed values with increasing density, and that density only determines a linear broadening in the spectra. Notably, the maximum real index saturates to a ``real-life'' value of $\sim 1.7$, in contrast to the indefinite growth predicted by both MB and LL.

We note that a number of experiments involving dense cold atomic clouds have observed both a saturation of the index \cite{Jenkins2016CollectiveExperiment,Jennewein2016CoherentCloud,Corman2017TransmissionAtoms} and the emergence of an anomalous broadening of the linewidth \cite{Pellegrino2014ObservationEnsemble,Jenkins2016CollectiveExperiment,Jenkins2016OpticalGases,Jennewein2016PropagationAtoms,Jennewein2018CoherentTheory}, including a linear scaling with density \cite{Jennewein2016CoherentCloud,Corman2017TransmissionAtoms}. 
A maximum index of $n\approx 1.26$ has also been observed in experiments involving dense, hot atomic vapours \cite{Keaveney2012MaximalNanolayer}, which has been attributed to atomic collisions. However, while complex collision dynamics necessitate semi-phenomenological models \cite{Allard1982TheLines}, here, our mechanism for saturation is quite fundamental, and occurs even for perfectly identical, stationary atoms.
%
%
%
%
%
%
%
\section{Introduction of RG scheme and refractive index analysis}
\label{sec:I.RG.S.&.R.I.A.}
Our RG theory is based upon the key intuition gained in the collective scattering of just two atoms, to build up an understanding of the many-atom problem in a hierarchical manner. To be specific, let us consider the problem of two identical atoms, whose distance is much smaller than a wavelength, $\rho_{12} \equiv k_0 |\vec r_1 - \vec r_2|\ll 1$. Applying \eqref{eq:CDS_input-output_eq_single_mode} and \eqref{eq:CDS_refractive_index_def_from_t}, we can calculate the imaginary part of the ``index" of the two-atom system, as illustrated in \figref{fig:RG_plots}-a. One can see that the characteristic two-atom spectrum (blue line) is not twice the response of a single, isolated atom (green dashed curve), but instead consists of two, well-separated peaks with different linewidths and shifted resonances. 

To understand this behavior, we consider the normal modes of the two-atom system, as encoded in the eigenstates of the dimensionless matrix $G$, whose elements \latestChanges{$G_{ij}$} were introduced in \eqref{eq:CDS_input-output_eq_single_mode2}. 
When $\rho_{12}\ll 1$, $G$ is dominated by its off-diagonal components $G_{12}=G_{21}$, and in particular, by the purely real $1/\rho_{12}^3$ near-field term (which we denote by $G_{12}^{\text{near}}$). Specifically, in spherical coordinates $\boldsymbol \rho_{ij} \equiv \rho_{ij} \left(\cos \theta\; \vhat x + \sin \theta \cos \phi\; \vhat y + \sin \theta \sin \phi\; \vhat z\right)$, one obtains $G_{ij}^{\text{near}}= 3(-1+3\cos^2 \theta)/(4\rho_{ij}^3) $. This describes the strong, coherent, near-field coupling between the two dipoles. This produces symmetric and anti-symmetric eigenstates whose dimensionless normal mode frequencies (real parts of the eigenvalues) are shifted as $\omega_{\pm}\approx \mp G_{12}^{\text{near}}$,
and align with the resonant peaks seen in \figref{fig:RG_plots}-a. \latestChanges{Given that $\im G$ is also a $2\times 2$ matrix with equal diagonal entries and equal off-diagonal entries, its eigenstates are also the same symmetric and anti-symmetric modes. This results in renormalized linewidths for these modes (given by the eigenvalues of $\im G$) of $\Gamma_{+}\approx 2$ and $\Gamma_{-}\approx \rho_{12}^2$, which is simply the two-atom limit of the famous Dicke superradiance model \cite{Gross1982Superradiance:Emission}. The key insight is that due to the large splitting, the total response in \figref{fig:RG_plots}-a is characterized by two well-separated resonances, which, although arising from the strong interaction of identical atoms, resemble the case of two, inhomogeneous and \textit{non-interacting} atoms, which were assigned these resonance frequencies and linewidths to start. This concept is at the heart of the RG approach for the many-atom case.}

We now discuss how strong, coherent $1/\rho_{ij}^3$ near-field interactions in a many-atom system can be treated, by successively replacing strongly interacting pairs by optically equivalent, non-interacting atoms. Here, we will focus on the main conceptual steps of our RG scheme, while additional justification of this scheme can be found in Sec. \ref{sec:M.J.RG.S.}. 
\latestChanges{Given the discussion above, we anticipate that the scheme generates an optically equivalent ensemble containing atoms with different renormalized resonance frequencies $\omega_i$. Contrary to the two-atom case, however, the linewidths will \textit{not} be renormalized within our RG scheme (see Sec. \ref{sec:M.J.RG.S.}). At any step of the RG flow, each pair of atoms} can either interact, or not, through the near-field coupling, depending on the previous RG steps. The normal modes of such a system are given by the eigenstates of the generalized $N \times N$ matrix $\mathcal M= \text{diag}( \boldsymbol \omega )-\tilde G$, where the elements $\tilde G_{ij}$ are defined as $\tilde G_{ij}= \mathcal L_{ij} G_{ij}^{\text{near}} + (G_{ij}-G_{ij}^{\text{near}})$. Here, $\text{diag}( \boldsymbol \omega )$ is a diagonal matrix containing the individual resonance frequencies $\boldsymbol \omega =(\omega_1,\dots,\omega_N)$, while $\mathcal L_{ij}=1$ or $0$ dictates whether pair $i,j$ is allowed to interact via the near field. \latestChanges{At the beginning of the RG process, the optically equivalent ensemble corresponds to the physical one, and thus all atoms are allowed to interact ($\mathcal L_{ij}=1$ for all pairs) and $\omega_i=\omega_0$. In three dimensions, the $1/\rho^3$ scaling of the near-field interaction implies that if an atom has a particularly close-by and near-resonant neighbour,} this pair will interact much more strongly between themselves than with any other nearby atoms \cite{Levitov1990DelocalizationInteraction}. Suppose that atoms $i,j$ (with $\mathcal L_{ij}=1$) are identified as the most strongly interacting pair, by a prescription given below. Then, we can re-write $\mathcal M$ as $\mathcal M=\mathcal M_{\text{pair}}+(\mathcal M-\mathcal M_{\text{pair}})$, where the only non-zero elements of $\mathcal M_{\text{pair}}$ involve atoms $i,j$. This effective $2\times 2$ matrix reads
\eq{
\label{eq:RG1_Mpair}
\mathcal M_{\text{pair}} =\mean{\omega}_{ij} \mathbb{I}
+
\left(\begin{array}{cc}
\delta\omega_{ij} & - G^{\text{near}}_{ij}\\\\
-  G^{\text{near}}_{ij} & -\delta\omega_{ij}
\end{array}
\right),}
where $\mean{\omega}_{ij}= (\omega_i +\omega_j)/2$ and $\delta\omega_{ij}= (\omega_i -\omega_j)/2$, and where we have included the coherent near-field interaction in $\mathcal M_{\text{pair}}$. The remaining far-field interactions between atoms $i$ and $j$, \latestChanges{as well as near- and far-field interactions involving all other atoms}, are included in $(\mathcal M-\mathcal M_{\text{pair}})$. The large near-field interaction motivates diagonalizing $\mathcal M_{\text{pair}}$ first, while treating $(\mathcal M-\mathcal M_{\text{pair}})$ as a perturbation. 

From the structure of $\mathcal M_{\text{pair}}$, we define the pairwise interaction parameter $\mathcal K_{ij}=\mathcal L_{ij}|G^{\text{near}}_{ij}|/(|\delta \omega_{ij}|+1)$. A large value of $\mathcal K_{ij}$ (which requires $\mathcal L_{ij}=1$) implies that the strong near-field interaction is able to strongly split the original resonances, including overcoming any possible differences in resonance frequencies $\delta \omega_{ij}$ of the pair. We thus identify the most strongly interacting pair as that with the largest value of $\mathcal K_{ij}$, as pictorially depicted in the first panel of \figref{fig:RG_plots}-b. Diagonalization of $\mathcal M_{\text{pair}}$ results in two, new interacting resonance frequencies $\omega_{\pm}=\mean{\omega}_{ij}\mp \sqrt{\delta\omega_{ij}^2 + (G_{ij}^{\text{near}})^2}$. We can then obtain an approximately equivalent system by replacing the two original resonance frequencies $\omega_{i,j}$ with the new values $\omega_{\pm}$ (second panel of \figref{fig:RG_plots}-b). While the resulting normal modes are in principle delocalized between atoms $i,j$, to facilitate the RG, we randomly assign $\omega_{+}$ to either atom $i$ or $j$, while $\omega_{-}$ is then assigned to the other atom (see Appendix C on the issue of replacing atoms $i,j$ with two new atoms placed at the midpoint of the original locations). 
This new system is described by a renormalized interaction matrix $\mathcal M_{\text{eff}}=\text{diag}(\boldsymbol{\omega}_{\text{eff}})-\tilde G_{\text{eff}}$, where $\boldsymbol{\omega}_{\text{eff}}=(\omega_1,\dots,\omega_{+}, \dots, \omega_{-},\dots \omega_N)$ contains the two renormalized resonance frequencies, and where $\tilde G_{\text{eff}}$ includes the new set of allowed near-field interactions $\mathcal L_{\text{eff}}$, which both forbid the renormalized pair from interacting again (i.e. $\mathcal L^{\text{eff}}_{ij}=0$) and prevent any backflow of the RG process (see Appendix B for more details). The RG process can be iteratively repeated by identifying, at each step, the most strongly interacting pairs, and ends once $\mathcal K_{ij} \leq \mathcal K_{\text{cut-off}}\sim 1$, i.e. when all strong near-field interactions have been removed. In the numerics presented here, we take a cutoff parameter of $\mathcal K_{\text{cut-off}}= 1$. Other choices result in minor quantitative corrections, while the overall conclusions remain the same. The final result, as suggested in the third panel of \figref{fig:RG_plots}-b, is that the original, homogeneous system can be mapped to an optically equivalent system that is inhomogeneously broadened, with a smooth probability distribution of resonance frequencies $P(\omega_{\text{eff}})$.

To validate the RG approach, we can use \eqref{eq:CDS_input-output_eq_single_mode} and \eqref{eq:CDS_input-output_eq_single_mode2} (with the near-field interactions of renormalized atoms suitably removed, see Appendix B) to calculate the maximum real index \latestChanges{(optimized over detunings)} as a function of density $\eta$ of the ensemble with renormalized resonance frequencies. This is plotted in \figref{fig:RG_plots}-c (green), along with exact numerical simulations (blue) of \eqref{eq:CDS_input-output_eq_single_mode} for the original system of identical atoms. These curves show good agreement for all densities, and in particular, reveal a maximum index of $n\approx 1.7$ at high densities. For comparison, the maximum index of the MB and LL equations (orange) increase indefinitely with density.

Furthermore, motivated by our previous observation that high-density spectra collapse onto the same curve when the detuning is rescaled by density (insets of \figref{fig:SM_spectrum} and \figref{fig:RG_plots}-c), in \figref{fig:RG_plots}-d, we plot the re-scaled probability distribution of effective resonance frequencies $P(\omega_{\text{eff}}/\eta)$ predicted by RG. For all densities considered ($2.5 \lesssim \eta \lesssim 80$), we see that a single universal curve results, i.e. the amount of broadening grows directly with density. Based on this curve, we find that the number of near-resonant atoms per reduced cubic wavelength $(\lambda_0/2\pi)^3=k_0^{-3}$, within a range $\pm\Gamma_0$ of the original atomic resonance frequency, is approximately $\sim 0.3$. The limited number of near-resonant atoms for light to interact with, regardless of how high the physical density is, directly explains the saturation of the maximum achievable index. We note that obtaining $P(\omega_{\text{eff}})$ by RG does not require solving the coupled equations of \eqref{eq:CDS_input-output_eq_single_mode2}, \latestChanges{but only the diagonalization of $2\times 2$ pairwise matrices, and we can calculate this distribution for much higher densities up to $\eta\sim 80$. Furthermore, as RG only involves the ``short-range'' near-field interaction (see \eqref{eq:RG1_Mpair}), we expect the rescaled distribution to be unique in the bulk of the atomic medium. That is, it should not depend sensitively on the specific geometry, provided that the system is sufficiently large that boundary effects are negligible. In \figref{fig:RG_plots}-d, the curves for $\eta\leq 3$ are obtained by a cylindrical geometry (the highest densities that we can compare to full coupled-dipole simulations, as in \figref{fig:RG_plots}-c). For higher densities $\eta>3$, when comparing with coupled-dipole simulations is no longer feasible, the extreme aspect ratio of the cylindrical geometry makes it inefficient to explore significantly higher densities using RG. We then find it more efficient to switch to atoms within a spherical geometry, which has the smallest surface area to volume ratio.}

Within the language of RG, the universal distribution $P(\omega_{\text{eff}}/\eta)$ constitutes the (numerically obtained) fixed point, as the interaction parameter of a system flows toward $\mathcal K_{ij}\rightarrow 1$. While it might be desirable to write down and analytically solve the RG flow equation for $P(\omega_{\text{eff}})$, this appears quite challenging in our case. This is because $\mathcal K_{ij}$ not only depends on the distance between atoms, but also their spatial orientation (as the near field is anisotropic) and the difference in resonance frequencies.

As mentioned earlier, it is rather inconvenient to derive key optical properties of a system, like index, by solving a set of equations (\eqref{eq:CDS_input-output_eq_single_mode2}) as large as the number of particles. At the same time, the RG approach clearly shows why conventional models (such as MB and LL) that treat atoms as a smooth medium fail at high densities \cite{Javanainen2014ShiftsSample,Javanainen2016LightOptics,Dobbertin2020CollectiveNanocavities}, \latestChanges{since the optical properties depend highly on granularity and on the strong interaction between an atom and a single, particularly close-by neighbor. Interestingly, RG also provides a basis to develop a more accurate smooth medium model. In particular, after the system is mapped to an inhomogeneously broadened distribution, $P(\omega_{\text{eff}})$, where near-field interactions and the influence of single neighbors are seen to be strongly reduced, one can finally apply a smooth medium approximation. Specifically, the MB equation (i.e. \eqref{eq:FTMS_index_prediction_MB}) for index can be readily generalized to an inhomogeneously broadened ensemble}
\eq{
\label{eq:FTMS_index_prediction_MB_RG}
n(\Delta)=\sqrt{1+3\pi \eta \Int \dfrac{ P(\omega_{\text{eff}})}{-\Delta + \omega_{\text{eff} }- i/2} d\omega_{\text{eff}}}.}
Substituting the distribution found in \figref{fig:RG_plots}-d, at high densities $\eta \gg 1$, this equation predicts a maximum index of $n\approx 1.8$, in good agreement with full results. \latestChanges{We stress that the emergence of a finite bound to the maximum index predicted by \eqref{eq:FTMS_index_prediction_MB_RG} can be directly related to the invariance of the distribution $P(\omega_{\text{eff}}/\eta)$, and thus to the linear growth of broadening with density.}
%
%

\latestChanges{

\begin{figure}[t!]
\centering
\includegraphics[width=1.\columnwidth]{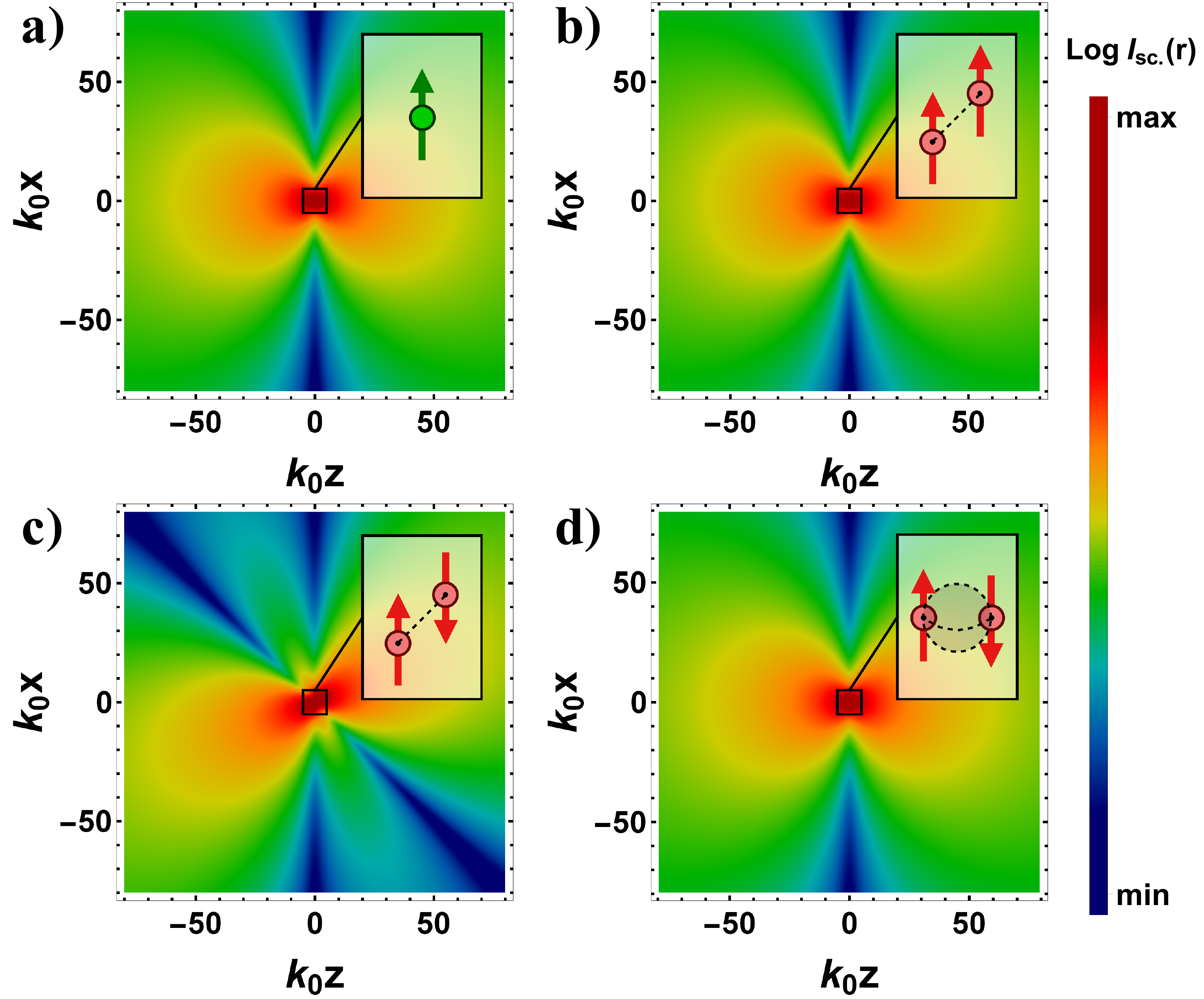}
\caption{
\latestChanges{
\textbf{Radiation pattern of a single dipole, compared to that of two in-phase or out-of-phase dipoles.} a) Given an isolated dipole $\vec d$ of fixed dipole amplitude $d_0$ and direction $\vhat x$, which is placed at $\vecRho=0$ and radiates light at the frequency $\omega_0$, we plot the intensity of the radiated field $I_{\text{sc.}}(\vec r)=|\mu_0 \omega_0^2\G(\vec r,\omega_0)\cdot \vec d |^2/(2\mu_0 c)$ in the $\vhat x$-$\vhat z$ plane, with the value indicated in the colorbar. b-c) Radiation pattern $I_{\text{sc.}}(\vec r)=|\mu_0 \omega_0^2\sum_{j=1,2}\G(\vec r - \vec r_j,\omega_0)\cdot \vec d_j |^2/(2\mu_0 c)$ for two near-positioned dipoles of fixed amplitude $d_0$ and direction $\vhat x$, oscillating either in-phase (b) or out-of-phase (c) with one another (i.e. $\vec d_1 = \pm \vec d_2 = d_0 \vhat x$), and placed at positions $\vecRho_1=-\vecRho_2=0.1(\vhat x + \vhat z)/\sqrt{2}$. d) Intensity radiated by two out-of-phase dipoles averaged over all possible inter-atomic orientations, keeping fixed the mutual distance $|\vecRho_1-\vecRho_2|=0.2$. This pattern closely resembles that of a single oscillating dipole.
}
}
\label{fig:RG_two_dipoles_p1}
\end{figure}

\section{Microscopic justification of the RG scheme}
\label{sec:M.J.RG.S.}
In the previous section, we have established that our RG procedure reproduces well the dependence of refractive index on density. We now present additional numerical and physical arguments that justify this approach and its approximations. Casual readers can consider skipping this section and jumping to Sec. \ref{sec:conclusions}. In this section, we will specifically answer the following questions:

\begin{enumerate}[label=\Alph*.]
\item Strictly speaking, RG is an approximate diagonalization of the many-atom near-field interaction matrix $G_{\text{near}}$, in terms of pairwise blocks. In sight of that, how well does our RG prescription reproduce the entire eigenvalue distribution of $G_{\text{near}}$?
\item In our RG prescription, the collective symmetric and anti-symmetric modes of two strongly interacting atoms are replaced by two new effective atoms with electric dipole transitions and modified resonance frequencies. However, this seemingly ignores the possibility that these modes (in particular the anti-symmetric one) could have a higher order multipolar character. Then, why is such a replacement valid?
\item As a related point, these collective modes can be renormalized again if they strongly interact with a third nearby atom. As a higher order multipolar mode can have a different scaling of the near field ($\sim 1/\rho^4$ in the case of the anti-symmetric mode), why does our replacement scheme with an electric dipole transition and a $\sim 1/\rho^3$ near-field interaction work?
\item Our RG prescription focuses on the strong interaction between nearby pairs due to the near field, but the $\sim 1/\rho$ far field associated with a radiating dipole might suggest that the large number of atoms \textit{far away} from a given atom might have a dominant effect in the interactions. What justifies treating the near field first over the far field?
\item As seen in the case of just two interacting atoms (\figref{fig:RG_plots}-a), both the resonance frequencies and linewidths of the collective modes are modified. Thus, why is it incorrect to renormalize both resonance frequencies and linewidths pairwise in the many-atom problem?
\end{enumerate}

\subsection{Comparison of eigenvalue distributions}
First, while we have previously focused on the observable quantity of refractive index, we note that mathematically, the RG approach is an attempt to approximately diagonalize the many-atom, near-field interaction matrix $G^{\text{near}}_{ij}= 3(-1+3\cos^2 \theta)/(4\rho_{ij}^3) $, in terms of pairwise blocks. We can thus test its accuracy by comparing the probability distribution of effective resonance frequencies $P(\omega_{\text{eff}} )$ obtained by RG, with the probability distribution of the eigenvalues of $-G_{\text{near}}$ obtained by exact diagonalization of a many-atom, dense system. A remarkable agreement can be observed in \figref{fig:RG_plots}-d, where the rescaled distribution of effective resonances $P(\omega_{\text{eff}} / \eta)$ is compared with the eigenvalue distribution of $-G_{\text{near}}$ (also rescaled by the density, black dashed curve), as calculated for the highest feasible density $\eta\approx 3$ of our cylindrical system. We separately checked that different (higher) densities and different geometries give similar results. Although subtle, we point out for future work the presence of a slight asymmetry in the exact eigenvalue spectrum around $\omega_{\text{eff}}=0$, which doesn't appear in the RG-derived distribution $P(\omega_{\text{eff}} )$. This might arise from higher order corrections to RG (e.g. rare triplets of nearly equidistant atoms, where the pairwise picture fails).

\subsection{Multipolar nature of collective modes}

Even if RG accurately predicts the resonance frequencies of a strongly interacting pair (e.g., the positions of the resonant peaks in \figref{fig:RG_plots}-a), one can wonder what is the justification of associating these two collective modes with two new individual atoms, which we implicitly assumed up to now to be characterized by electric dipole transitions like the original atoms. 

To frame the issue, we recall from Sec. \ref{sec:I.RG.S.&.R.I.A.} that two strongly interacting, identical atoms are diagonalized by a symmetric and an anti-symmetric collective mode, where the two atomic electric dipoles respectively oscillate in phase or out of phase with one another. Clearly, the symmetric mode retains an electric dipole character, as the two individual dipoles add to produce a dipole of doubled amplitude. This can be observed in \figref{fig:RG_two_dipoles_p1}, where we compare the intensity pattern radiated by one single dipole $\vec d=d_0\vhat x$ of fixed amplitude and direction (\figref{fig:RG_two_dipoles_p1}-a) with that of two in-phase, close-by dipoles $\vec d_1=\vec d_2=d_0\vhat x$ (\figref{fig:RG_two_dipoles_p1}-b). 
The case of the anti-symmetric mode, however, is visibly more complex (\figref{fig:RG_two_dipoles_p1}-c). Intuitively, the two out-of-phase dipoles produce a vanishing electric dipole response, and are instead a hybrid of magnetic dipole and electric quadrupole modes. Interestingly, though, while the radiation pattern of \figref{fig:RG_two_dipoles_p1}-c depends sensitively on the relative orientation of the two out-of-phase dipoles, if one averages over orientations, the pattern again closely resembles that of a single, electric dipole (see \figref{fig:RG_two_dipoles_p1}-d). More concretely, in Appendix D we show that the orientation-averaged resonant scattering cross section associated to the anti-symmetric mode is $\mean{\sigma_{\text{sc}}^-} \approx 0.94 \sigma_{\text{sc}}$, where we recall that $\sigma_{\text{sc}}=3\lambda_0^2/(2\pi)$ is the resonant cross section of a single atom with electric dipolar response. To sum up, the anti-symmetric mode on average is seen to behave almost identically to a single atom with electric dipole response, justifying such a replacement in our RG prescription. Furthermore, we show in Appendix D that this agreement is even stronger when considering pairs of strongly interacting atoms with different resonance frequencies $\delta\omega_{ij}\neq 0$, which is a situation typically encountered in an actual RG flow.

\begin{figure}[t!]
\centering
\includegraphics[width=\columnwidth]{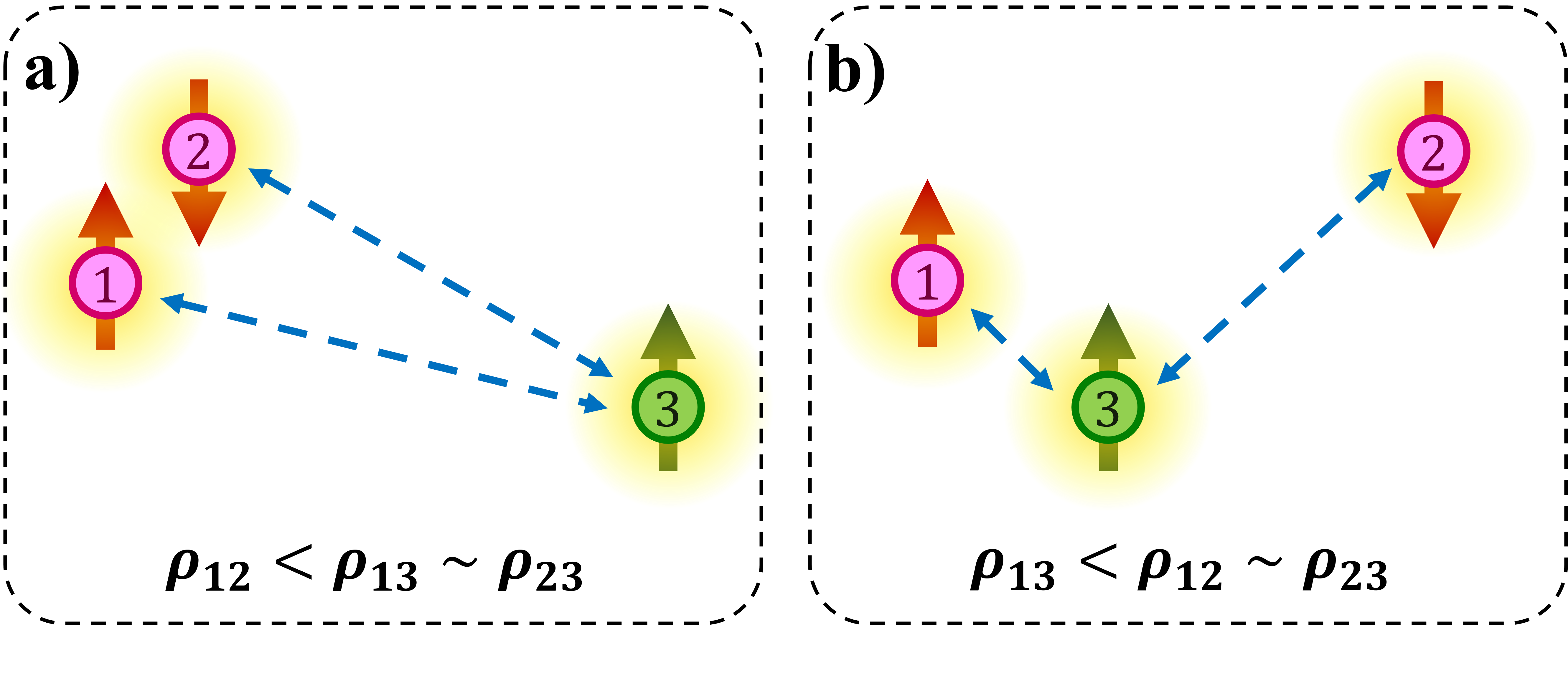}
\includegraphics[width=0.94\columnwidth]{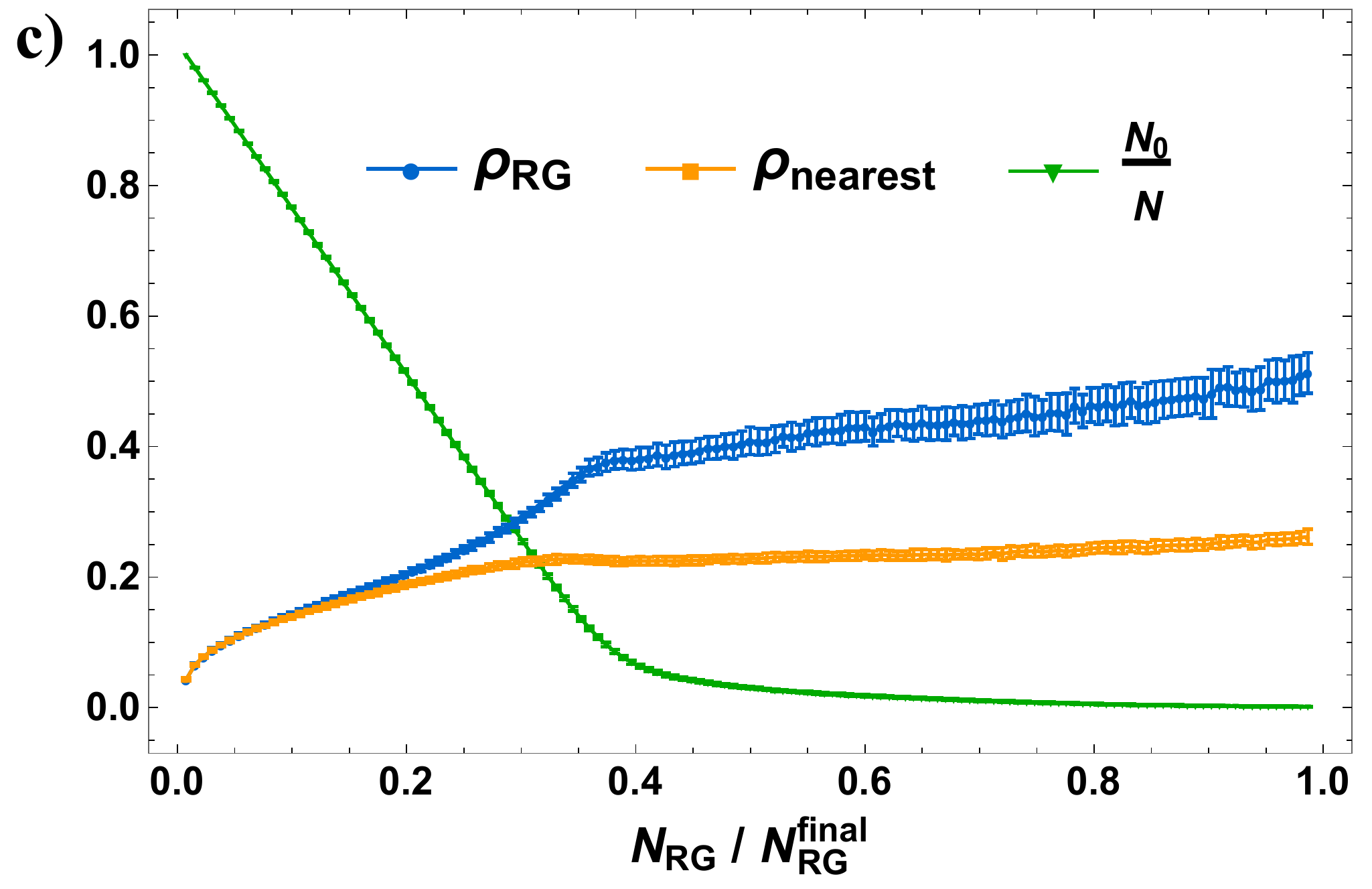}
\caption{
\latestChanges{
\textbf{Microscopic analysis of the renormalization of the anti-symmetric modes.} 
a) Pictorial representation of the near-field interaction between the anti-symmetric mode of a pair (represented by two out-of-phase dipoles, pink circles labeled $1$ and $2$) and a third atom (green circle, with label $3$), which may sit very far from the pair (characterized by the distances $\rho_{12}< \rho_{13}\sim \rho_{23}$). In this case, the interaction strength scales like $\sim 1/\rho_{13}^4$, reflecting the higher order multipole nature of the out-of-phase dipoles. b) A similar illustration for the case where the third atom sits closer to one atom of the pair (say atom 1) than the pair separation itself ($\rho_{13}<\rho_{12}\sim \rho_{23}$). The interaction strength then scales like $\sim 1/\rho_{13}^3$.
c) System properties during the RG flow. The horizontal axis quantifies how many pairs $N_{\text{RG}}$ have been renormalized, from the beginning ($N_{\text{RG}}=0$) towards the end ($N_{\text{RG}}=N_{\text{RG}}^{\text{final}}$) of the algorithm. As one atom can be renormalized more than once, typically $N_{\text{RG}}^{\text{final}}>N/2$. The blue circles represent the average inter-atomic distance $\rho_{\text{RG}}$ of those pairs that get renormalized, while the orange squares shows the average distance $\rho_{\text{nearest}}$ between the atoms of those pairs and their respective nearest atom, chosen among those that are still allowed to interact (i.e. with $\mathcal L_{ij}=1$). %
The green triangles display the fraction of atoms $N_0/N$ that have never been renormalized up to that moment of the flow. The data represents the average over $\sim 300$ runs over different random atomic positions, uniformly sampled inside a sphere of radius $r_{\text{sph.}}=0.55\lambda_0$ and density $\eta \approx 32$. The bars show one standard deviation in the accumulated statistics.
}
}
\label{fig:RG_distance}
\end{figure}

\subsection{Near-field interaction involving renormalized atoms}

Having argued that the anti-symmetric mode has an average optical response resembling that of a single electric dipole, we now turn to a second, related issue. Namely, since the anti-symmetric mode is a hybrid of magnetic dipole and electric quadrupole modes, it should have a near-field of $\sim 1/\rho^4$ at distances $\rho$ much larger than the separation between the two composing atoms and much smaller than the optical wavelength. As the RG flow proceeds, a third atom that interacts strongly with this mode would then see such a scaling law at this distance (see \figref{fig:RG_distance}-a). However, our RG prescription assumes that any new effective resonance has electric dipole character, and in particular, a $\sim 1/\rho^3$ near-field interaction with the third atom. We now argue that the RG prescription is a good approximation, because as the RG flow continues, it is likely that the third atom actually sits \textit{closer} to one of the atoms in the pair (say atom 1), than the pair separation itself (\figref{fig:RG_distance}-b). In that case, the effective interaction strength between the anti-symmetric mode of the pair and the third atom will scale as $\sim 1/\rho_{13}^3$, exactly as if this mode was replaced by an electric dipolar atom. Mathematically, this is possible because the interaction parameter $\mathcal K_{ij}=|G^{\text{near}}_{ij}|/(|\delta \omega_{ij}|+1)$ that governs when atoms are renormalized does not depend only on closest distance of separation (via $G^{\text{near}}_{ij}$), but on the detunings $\delta \omega_{ij}$ as well. To quantify this picture, we have run $\sim 300$ RG flows over random configurations of a dense medium ($\eta=32$) within a spherical geometry of radius $r_{\text{sph.}}=0.55\lambda_0$. In \figref{fig:RG_distance}-c, we plot several salient properties throughout the RG flow, averaged over the various runs. The horizontal axis denotes the relative position within the flow ($0\leq N_{\text{RG}}/N_{\text{RG}}^{\text{final}} \leq 1$). In particular, $N_{\text{RG}}^{\text{final}}$ is the total number of pairs renormalized during the entire RG (starting from a homogeneous atomic medium, until one reaches $\mathcal K_{ij}< \mathcal K_{\text{cut-off}}=1$ for all pairs), while $N_{\text{RG}}$ denotes the total number of renormalized pairs at any point in between. We recall that it is possible for an atom to be renormalized more than once, so that in general $N_{\text{RG}}^{\text{final}}>N/2$ for a dense medium. For reference, in green, we plot the fraction $N_0/N$ of atoms that have never been renormalized up to that point. Notably, the fact that $N_0/N$ reaches nearly zero when $N_{\text{RG}}/N_{\text{RG}}^{\text{final}} \sim 0.4$ indicates that almost all renormalization events beyond this stage involve previously renormalized (and thus inhomogeneous) atoms. Separately, with blue circles, we show the average value of the inter-atomic distance between atoms comprising the renormalized pairs at that stage, and we compare it with the average distance between each atom of these pairs and its own nearest neighbour, as portrayed by the orange squares. As we are interested in the interaction between atoms which will possibly be renormalized in some subsequent RG step, we only count the nearest neighbours where $\mathcal L_{ij}=1$. The figure shows that we can roughly divide the RG flow into two parts. Before the critical value of $N_{\text{RG}}/N_{\text{RG}}^{\text{final}} \sim 0.4$, many atoms are still homogeneous, so that the algorithm mostly renormalizes pairs of identical, nearest neighbour atoms (as confirmed by the coincidence of the blue and orange curves). On the contrary, when $N_{\text{RG}}/N_{\text{RG}}^{\text{final}} \gtrsim 0.4$, almost all atoms have already been renormalized at least once, and in particular, an effective atom representing an anti-symmetric mode can potentially be renormalized again. In this regime, however, the nearest, interacting neighbour to the two original atoms forming this mode is on average significantly closer than the distance between these two atoms, as evidenced by the blue curve being significantly higher than the orange. This confirms that the intuitive picture of \figref{fig:RG_distance}-b constitutes a typical case, which preserves the $\sim 1/\rho^3$ scaling of the near-field interaction.


%
%

\subsection{Near-field vs. far-field interactions}

Separately, we want to underline the importance of separating the effects of near-field and far-field interactions, which occur in an atomic medium.
To this aim, we point out the historic work of \cite{Levitov1990DelocalizationInteraction}, which used RG to understand the properties of permanent, \textit{static} dipoles, which only experience a near-field $1/\rho^3$ interaction. Given only a near-field interaction in three dimensions, the interaction of a dipole with its nearest neighbour is then indeed dominant. However, we have a qualitatively different system, of driven, radiating dipoles. Naively then, a similar argument considering the $1/\rho$ far field would suggest that atoms within a shell of radius $\rho$ and $\rho+d\rho$ of one atom at the origin would contribute an interaction strength of $\sim \rho d\rho$, such that the \textit{furthest} atoms actually play the strong role. We argue that an RG process based on the near field is still the correct prescription, as the index should be a local property. Instead, the apparent "dominance" of the far field simply reflects the fact that the macroscopic geometry of an optical system (e.g., if it is shaped as a lens or prism) can drastically alter the overall optical response, but not the index.

\subsection{Linewidths in the RG prescription}

Finally, we note that although the problem of just two atoms (\figref{fig:RG_plots}-b) can be interpreted in terms of renormalized resonance frequencies and linewidths, in the many-atom case, we \textit{only} renormalize the resonance frequencies. As we discussed, the interaction between atoms is described by the dimensionless matrix $G$ (as defined in \eqref{eq:CDS_input-output_eq_single_mode2}), whose real part $\re G$ determines the coherent part of the interaction (i.e. the collective resonance frequencies), while its imaginary part $\im G$ is associated to the dissipative phenomena, thus dictating the collective linewidths. In the case of two identical atoms, $\re G$ and $\im G$ are both naturally and exactly diagonalized by the same symmetric and anti-symmetric modes. 
However, in a many-atom ensemble, the different mathematical structures and physical origins of $\re G$ and $\im G$ become important. In particular, we recall that the $\sim 1/\rho_{ij}^3$ near-field component of the Green's function, $G_{ij}^{\text{near}}$, is purely real and strongly divergent as two atoms approach each other, which motivates our RG theory based on diagonalizing these terms first. Physically, $\im G$ does not contain a near-field term (recall that $\im G_{ij} \rightarrow 1/2$ as $\rho_{ij}\rightarrow 0$), since dissipation is associated with the radiation of energy into the far field. The absence of a near-field term implies that $\im G$ does not yield an especially strong interaction between close atomic pairs, and thus cannot be approximately diagonalized pairwise. Again, this makes sense physically, because the emitted power by a collection of dipoles depends on the global interference between all dipoles, and does not generally decompose into the sum of powers radiated by pairs.
Separately, we have checked that if our RG prescription were modified to renormalize resonance frequencies \textit{and} linewidths pairwise, it would predict a non-physical optical response that tends to decrease ($n\rightarrow 1$) in the limit of high densities, in contrast with the full numerical simulations.
}
%
%
%
%
%
%
%
\section{Conclusions}
\label{sec:conclusions}
To summarize, we have shown that despite the large resonant scattering cross section of a single atom, a dense atomic medium does not exhibit an anomalously large optical response. Rather, strong near-field interactions between atomic pairs combined with spatial disorder results in an effective inhomogeneous broadening mechanism, which occurs even if the atoms are otherwise perfect, and yields a maximum index of $n\approx 1.7$. The key role of atomic granularity in this process also illustrates why conventional smooth medium approximations fail to describe the near-resonant response.

While we have focused on the linear refractive index, we believe that our RG formalism is valid in general for resonant disordered atomic media, and constitutes a versatile new tool to study multiple scattering. Within the linear regime, RG might be used to provide additional insight to the question of whether an Anderson localization transition exists in a 3D ensemble, and under what conditions \cite{Skipetrov2014AbsenceScatterers,Bellando2014CooperativeHamiltonian,Skipetrov2016RedLocalization,Skipetrov2018Ioffe-RegelScatterers,Cottier2019MicroscopicLight,Skipetrov2019SearchField}. 
\latestChanges{Furthermore, it would be interesting to explore the usage of RG toward the challenging problem of quantum, nonlinear scattering. As previously mentioned, the multiple scattering problem is formally encoded in a non-Hermitian Hamiltonian that describes light-mediated dipole-dipole interactions between atoms. In the limit of linear response, the resulting equations are equivalent to our coupled-dipole equations of \eqref{eq:FTMS_coupled_dipoles}, but beyond that, one is faced with the challenge of dealing with the exponentially large Hilbert space associated with $N$ two-level atoms. Perturbative diagrammatic approaches have only recently been developed to treat the dilute atom limit \cite{Binninger2019NonlinearCloud}, but our understanding of the nonlinear physics beyond this regime is very limited. To this end, we hypothesize that a diagrammatic theory can also be developed in the dense, strong scattering regime, where strong interactions between nearby pairs are first non-perturbatively summed via RG, while remaining interactions can be treated perturbatively.}

Our results could also have interesting implications for quantum technologies based on atomic ensembles. In particular, the total optical depth of system, given by the product of the imaginary part of the index and system length, $D\sim (\im n)k_0 L$, is a fundamental resource \cite{Gorshkov2007UniversalMedia,Hammerer2010QuantumEnsembles,Gorshkov2011Photon-PhotonBlockade}, with its magnitude establishing fundamental error bounds for most applications. As the imaginary part of the index also saturates with increasing density, this could place minimum size constraints on systems in order to achieve a given fidelity. Likewise, constraints on the maximum density could arise due to the induced inhomogeneous broadening, which typically constitutes an undesirable dephasing mechanism. 

Finally, it would be interesting to understand more fully how the optical properties of a dilute atomic medium eventually transform into the low refractive index of actual optical materials, as the density is increased. Specifically, for a disordered ensemble, we have seen that the maximum index already saturates, at densities that are approximately six orders of magnitude before the onset of chemical processes. We hypothesize that the onset of chemistry, and the phase transition toward a real material, does not qualitatively alter the optical response, provided that the system remains disordered and the electrons tightly bound. Separately, it would be interesting to explore the same questions and transition for spatially ordered atomic systems, where RG breaks down and one expects very different qualitative behavior, due to the possibility of strong constructive and destructive interference in light scattering. \latestChanges{We note that there have been recent efforts to predict when a high index might occur within solid-state materials \cite{Naccarato2019SearchingStudy,Shubnic2020HighNanophotonics}, and it would be interesting in future studies to develop a full theory combining quantum chemistry and multiple scattering, to explore the transition from dilute atomic media to real materials.}
%
%
%
%
\section*{Acknowledgements}
We acknowledge S. Grava, N. Fayard, J.-J. Greffet, S. Wu, and H.J. Kimble for stimulating discussions. F.A. acknowledges support from the ICFOstepstone - PhD Programme funded by the European Union’s Horizon 2020 research and innovation programme under the Marie Skłodowska-Curie grant agreement No 713729. D.E.C. acknowledges support from MINECO Severo Ochoa Grant CEX2019-000910-S, Generalitat de Catalunya through the CERCA program, Fundació Privada Cellex, Fundació Mir-Puig, the European Union’s Horizon 2020 research and innovation programme, under European Research Council grant agreement No 639643, FET-Open grant agreement No 899275 (DAALI), and Quantum Flagship Project 820445 (QIA), Plan Nacional Grant ALIQS (funded by Ministerio de Ciencia, Innovacion y Universidades, Agencia Estatal de Investigacion, and European Regional Development Fund), Fundación Ramón Areces Project CODEC, the Europa Excelencia program funded by Agencia Estatal de Investigacion (project number EUR2020-112155, ENHANCE), and QuantumCAT (funded within the framework of the ERDF Operational Program of Catalonia, ref. 001-P-001644).

\setcounter{equation}{0}
\def\theequation{A.\arabic{equation}}

\setcounter{figure}{0}
\def\thefigure{A.\arabic{figure}}

\begin{figure*}[t!]
\centering
\includegraphics[width=2\columnwidth]{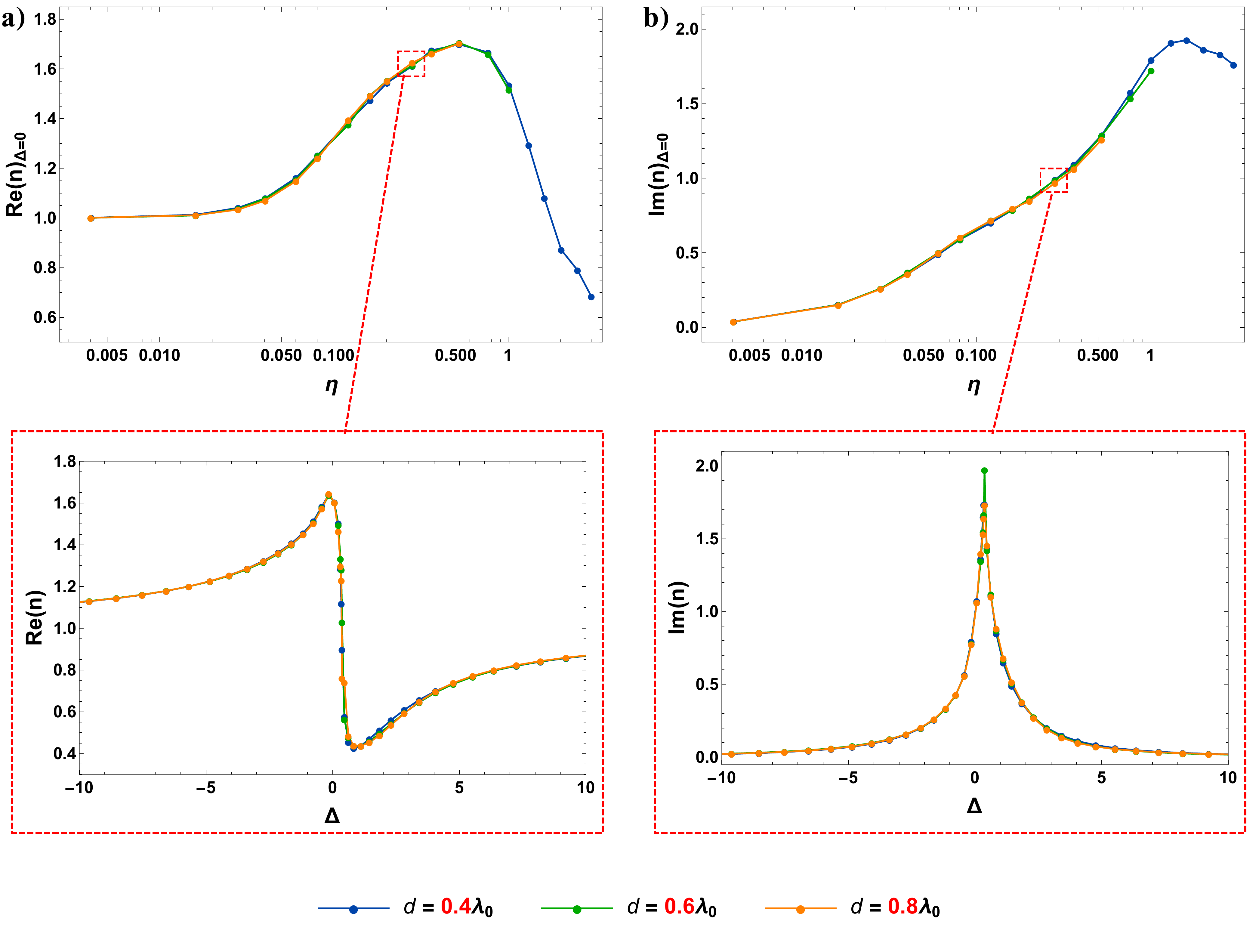}
\caption{\textbf{Independence of the refractive index from the thickness of the ensemble.} Given the physical system of Fig. 2-a of the main text (with $w_0=2.5\lambda_0$, $l_{\text{cyl.}}=5\lambda_0$), we compare the resonant ($\Delta = 0$) refractive index as a function of the density, for various ensemble thicknesses: $d=0.4\lambda_0$ (as in the main text, here in blue), $d=0.6\lambda_0$ (in green) and $d=0.8\lambda_0$ (in orange). Subfigures a) and b) illustrate the real and imaginary parts of the index, respectively. The insets show the full spectra $n(\Delta)$ at a fixed density $\eta \simeq 0.28$. All data are obtained by averaging $\mean{ t(\Delta)}$ over $>1000$ configurations.
} 
\label{fig:SI_thick_indep}
\end{figure*}

\section*{Appendix A: Linear behaviour of the index as a function of thickness}
Our operative definition of the complex index of refraction, as given by Eq. 7 of the main text, is
\eq{
\mean{t(\Delta)}= \exp \left\{ i  \left[n\left(\Delta \right)-1\right] k_0 d\right\}.
}
Since the refractive index is an intensive property by definition, it must not depend upon the thickness $d$ that we choose in our numerics. Here, we show that our operative definition satisfies this condition.

We consider the same physical system described in Fig. 2-a of the main text, with $w_0=2.5\lambda_0$, $l_{\text{cyl.}}=5\lambda_0$ and different values of the thickness. By applying Eq. 6 and Eq. 7 of the main text, we compute the resonant ($\Delta = 0$) refractive index for growing values of the density $\eta$, and we plot its real (imaginary) part in \figref{fig:SI_thick_indep}-a(b). The simulated values of the thickness are: $d=0.4\lambda_0$ (as in the main text, here in blue), $d=0.6\lambda_0$ (in green) and $d=0.8\lambda_0$ (in orange). Moreover, for the point at $\eta\simeq 0.28$, we evaluate the full spectra $n(\Delta)$, as represented in the insets of the figure. All curves show the same behaviour, independently of $d$, both on resonance and when varying the detuning.

\section*{Appendix B: Full description of the RG algorithm}
Here, we provide a full description of the RG algorithm. We assume that we have an ensemble of $N$ randomly positioned atoms. As shown in \eqref{eq:CDS_input-output_eq_single_mode2}, each pair of atoms interacts through the coupling $G_{ij}= ( 3\pi/k_0) \vhat x \cdot \G(\boldsymbol \rho_{ij},\omega_0)\cdot \vhat x$, where $\boldsymbol \rho_{ij} \equiv k_0 (\vec r_i-\vec r_j)$. The $1/\rho_{ij}^3$ near-field component of $G_{ij}$ reads
\eq{
\label{eq:AppB_Gnear_definition}
G_{ij}^{\text{near}}= \dfrac{3}{4\rho_{ij}^3}(-1+3\cos^2 \theta) ,
}
where we have represented $\boldsymbol \rho_{ij} \equiv \rho_{ij} (\cos \theta, \sin \theta \cos \phi, \sin \theta \sin \phi)$ in spherical coordinates. Here, we define $G_{jj}^{\text{near}}=0$, in accordance with the definition $G_{jj}=i/2$. This near-field interaction is purely real, and describes a coherent interaction between dipoles. 

Let us now consider a generic step of the RG flow, where the atomic ensemble is already composed of effective atoms characterized by different atomic resonances and a specific set of allowed near-field interactions. As discussed in the main text, this system is described by the $N\times N$ matrix $\mathcal M= \text{diag}( \boldsymbol \omega )-\tilde G$, where the elements $\tilde G_{ij}$ read $\tilde G_{ij}= \mathcal L_{ij} G_{ij}^{\text{near}} + (G_{ij}-G_{ij}^{\text{near}})$. Numerically, this matrix is initialized according to $\boldsymbol \omega ^{\text{init.}}=(0,\dots,0)$ and $\mathcal L_{ij}^{\text{init.}}=1-\delta_{ij}$, stating that all atoms are resonant at the frequency $\omega_0$ and cannot self-interact. 

At each step of the RG flow, we evaluate the list of couplings $\mathcal K_{ij}=\mathcal L_{ij}|G^{\text{near}}_{ij}|/(|\delta \omega_{ij}|+1)$ (where $\delta\omega_{ij}= (\omega_i -\omega_j)/2$), ordering them from the largest to smallest in amplitude. Nominally, we should select the most strongly interacting pair and renormalize the pair properties, but the computational cost of this approach would be unfeasible for large atom number. Due to this reason, we start from the most strongly interacting pair (say, $i,j$), select it, and remove from the list all other pairs containing one of those atoms (e.g. $i,k$ or $j,k$). We then proceed iteratively, until we select $N_{\text{step}}$ most strongly interacting pairs. We choose $N_{\text{step}}$ to be a small fraction of the total atom number $N$ (approximately $\sim 2.5\%$), since the maximum number of possible disjoint pairs scales as $N/2$. Nevertheless, we have checked that the results are insensitive to different choices. 

Given each pair $(i,j)$ of the selected set, we diagonalize $\mathcal M_{\text{pair}} $, and define its eigenvalues as the new effective resonances $\omega_{\pm}=\mean{\omega}_{ij}\mp \sqrt{\delta\omega_{ij}^2 + (G_{ij}^{\text{near}})^2}$, where $\mean{\omega}_{ij}= (\omega_i +\omega_j)/2$. We then substitute the initial frequencies ($\omega_i$, $\omega_j$) with the new two effective resonances in $\boldsymbol{\omega}$, the order of the labels being chosen randomly. 

We need to impose that the pair does not interact anymore through the near field, meaning that we must replace $\mathcal L_{ij}^{\text{old}}=1$ with $\mathcal L_{ij}^{\text{new}}=0$. At the same time, at any given stage of the RG flow, the resonance frequencies of any pair of effective atoms $i$ and $j$ might have been derived from a set of previous RG steps involving a set of atoms with indices $\{I'\}$ and $\{J'\}$, respectively. If the sets $\{I'\}$ and $\{J'\}$ have some non-zero intersection, then atoms $i$ and $j$ must be omitted from a subsequent frequency renormalization step. Not doing this would violate the principle of RG, that we are integrating or ``freezing" out the degrees of freedom with the strongest interactions. Numerically, we efficiently enforce this constraint by replacing $\mathcal L_{ik}^{\text{new}}=\mathcal L_{jk}^{\text{new}}=\mathcal L_{ki}^{\text{new}}=\mathcal L_{kj}^{\text{new}}=\mathcal L_{ik}^{\text{old}}\mathcal L_{jk}^{\text{old}},$ $\forall k$, anytime a pair $(i,j)$ is renormalized. Since $\mathcal L$ has (at any step) zero-valued diagonal elements, this directly ensures that $\mathcal L_{ij}^{\text{new}}=0$.

After all atoms of the step have been renormalized, we re-evaluate the new set of $\mathcal K$ parameters, and repeat the scheme. When all pairs exhibit $\mathcal K\leq \mathcal K_{\text{cut-off}}=1$, we stop the RG flow, obtaining an ensemble of $N$ inhomogeneously broadened atoms. Given a fixed value of the density $\eta$, we repeat this process for $\approx 100$ different spatial configurations, in order to build up the final distribution $P(\omega_{\text{eff}})$.

We extract the optical properties from the renormalized ensemble by applying \eqref{eq:CDS_input-output_eq_single_mode2} of the main text, modified in order to account for the the new $N\times N$ matrix $\mathcal{M}$ emerging from the RG scheme. This reads
\eqbreak{
\label{eq:Methods_input-output_eq_single_mode}
\left(-\Delta + \omega_i\right) c_i(\Delta)-\displaystyle
\Sum_{j=1}^{N} \left[G_{ij}-\left(1-\mathcal L_{ij}\right)G_{ij}^{\text{near}}\right] c_j(\Delta)  \\\\ = \; \dfrac{ E_{\text{in}}(\vec r_i,\omega_0)}{E_0}.}

%
%
%
%
%
%
%
%
%
%

\section*{Appendix C: Definition of effective positions in the RG scheme}
In the main text (cf. \figref{fig:RG_plots}-b), we described how the optical response of a pair of atoms separated by a distance $\rho_{ij}\ll 1$ is characterized by two effective resonance frequencies, corresponding to the real parts of the eigenvalues of the two-atom system. The two collective modes are intrinsically delocalized in space (being formed by atoms with two different positions $\vec r_{i,\;j}$). As this delocalization is difficult to incorporate into the RG scheme, we instead attribute each of these two resonance frequencies to a new effective atom, with well-defined position.

In the main text, it was stated that the new effective atomic positions are assigned to those of the original pair, $\vec r_{i,\;j}$ (randomly between the two possible permutations). A more natural choice, given that the two renormalized atoms are non-interacting, might be to place them at the midpoint  $(\vec r_{i}+\vec r_{j})/2$ between the two original atoms, but here we discuss the problem with that approach.

\begin{figure}[t!]
\centering
\includegraphics[width=\columnwidth]{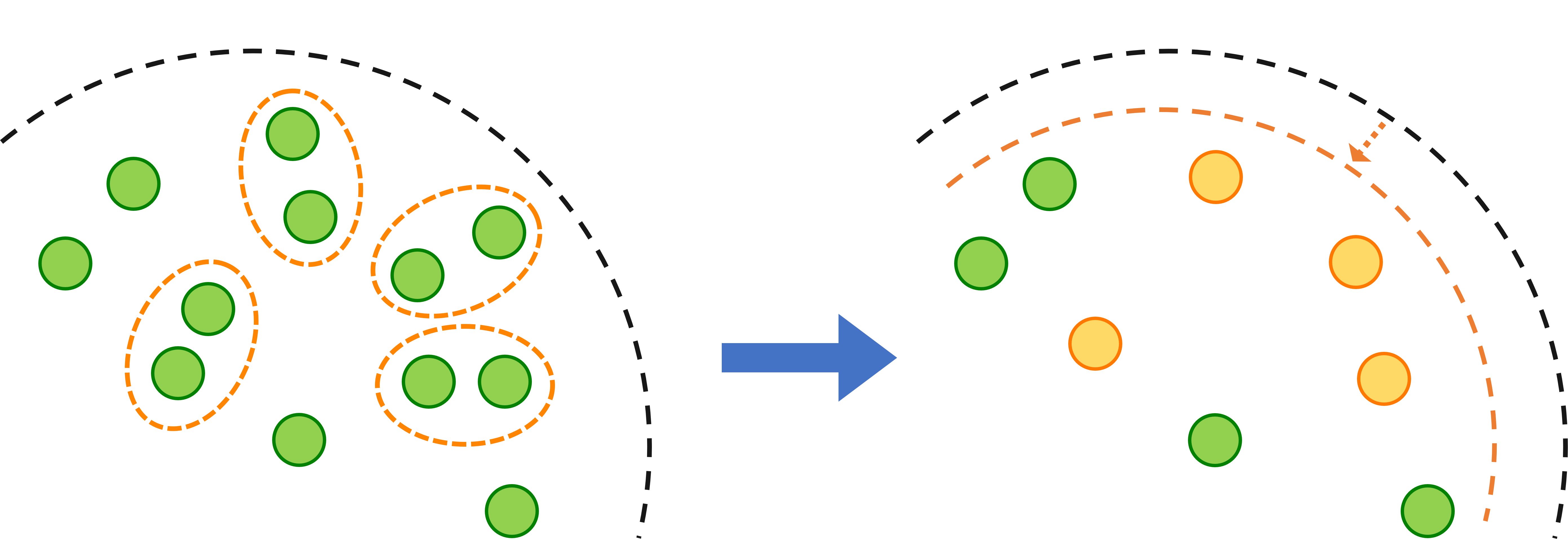}
\caption{\textbf{RG scheme based upon re-positioning atoms.} Due to the finite size of the sample, if one defines the positions of the new effective atoms as being at the midpoint between the original pair, then, at each RG step, the cloud effectively shrinks, resulting in a distortion of the ensemble.} 
\label{fig:SI_shrink}
\end{figure}

Specifically, for a finite-size sample, the atoms closest to the perimeter of the sample will only renormalize with atoms that are closer to the interior. As illustrated in \figref{fig:SI_shrink}, this means that step by step, the shape of the cloud tends to shrink. This effectively distorts the ensemble and results in a higher density, and higher interaction strengths in the next step of RG.

\latestChanges{
\section*{Appendix D: scattering cross section of two near-positioned atoms}
The optical response of an identical atomic pair is characterized by a symmetric and an anti-symmetric normal mode. Here, we will study the scattering cross sections of such modes, in the limit of near-positioned atoms. 

First of all, let us write the dimensionless positions (in units  of $k_0^{-1}$) of the two atoms of the pair as
\eq{
\boldsymbol \rho_1 = - \vecRho_2 = \dfrac{\rho_{12}}{2}\left(\cos \theta\; \vhat x + \sin \theta \cos \phi\; \vhat y + \sin \theta \sin \phi\; \vhat z\right),
}
where $\rho_{12}\ll 1$. The scattering cross section can be derived by means of the \textit{so-called} optical theorem \cite{Newton1976OpticalBeyond,Draine1988TheGrains,Jackson1998ClassicalElectrodynamics,Sosa2003OpticalShapes,Evlyukhin2011MultipoleApproximation,Evlyukhin2013MultipoleSurface,Alaee2020KerkerAntennasb}, which reads
\eq{
\label{eq:appE_optical_theorem}
\sigma_{\text{sc}}^{\text{pair}}(\Delta) = \dfrac{\sigma_{\text{sc}}}{2} \im \Sum_{j=1}^2 \dfrac{E_{\text{in}}^*(\vecRho_j)}{E_0} c_j(\Delta)
}
where $\sigma_{\text{sc}}=3\lambda_0^2/(2\pi)$ is the resonant cross section of a single, isolated, electric dipolar atom, while the dimensionless coefficients $c_j(\Delta)$ are defined as in \eqref{eq:CDS_input-output_eq_single_mode2} of the main text.

By plugging the solutions of \eqref{eq:CDS_input-output_eq_single_mode2} into \eqref{eq:appE_optical_theorem}, one obtains a total cross section characterized by the two resonances $\omega_{\pm}$, which are respectively associated to the symmetric and anti-symmetric modes, so that the resonant scattering cross sections of these two modes can be defined as $\sigma_{\text{sc}}^{\pm}\equiv \sigma_{\text{sc}}^{\text{pair}}(\omega_{\pm})$. In the limit where $\rho_{12}\ll 1$, the two resonances $\omega_{\pm}=\mp \re G_{12}\sim \mp 1/\rho_{12}^3$ are well-separated and can be efficiently resolved, leading to
\eq{
\label{eq:AppE_scatt_cross_plus_minus}
\dfrac{\sigma_{\text{sc}}^{\pm}}{\sigma_{\text{sc}}}\simeq \dfrac{(\vec E_{12}^{*}\cdot \vec v_\pm)( \vec E_{12}\cdot \vec v_\pm)}{\Gamma_\pm  },
}
where we defined $\vec E_{12}\equiv\{E_{\text{in}}(\boldsymbol \rho_1),E_{\text{in}}(\boldsymbol \rho_2)\}/E_0$, as well as the eigenstates $\vec v_\pm =\{1,\pm 1\}/\sqrt{2}$ and the decay rates $\Gamma_{\pm}=1\pm2\im G_{12}$.

Assuming that the input field is either a $\vhat x$-polarized, $\vhat z$-directed Gaussian beam with $w_0\gg\lambda_0$ and focal point at $\vec r =0$, or equivalently a $\vhat x$-polarized, $\vhat z$-directed a plane wave, one can evaluate the cross sections in the limit of $\rho_{12}\ll 1$, obtaining
\eq{
\label{eq:AppE_scatt_cross_sections_ratio}
\dfrac{\sigma_{\text{sc}}^+}{\sigma_{\text{sc}}}\simeq 1,\;\;\;\;\;
\dfrac{\sigma_{\text{sc}}^-(\theta,\phi)}{\sigma_{\text{sc}}}\simeq \dfrac{f(\theta,\phi)}{g(\theta)},
}
where $f(\theta,\phi)\equiv ( \sin \theta \sin \phi)^2$ and $g(\theta)\equiv [2-\cos^2\theta]/5$. 

As expected, the symmetric mode exhibits a perfect electric dipolar behaviour, characterized by the same scattering cross section of one single, isolated atom. On the contrary, the complex multipolar nature of the anti-symmetric mode leads to a more complicated scattering cross section, which depends on the mutual orientation of the initial pair. This suggests considering the average resonant cross section over all possible orientations of a pair, obtaining
\eq{
\left\langle\dfrac{\sigma_{\text{sc}}^-}{\sigma_{\text{sc}}}\right\rangle
=\dfrac{1}{4\pi}\Int_0^{\pi}  d\theta \Int_0^{2\pi}  d\phi
\;\dfrac{\sigma_{\text{sc}}^-(\theta,\phi)}{\sigma_{\text{sc}}}\sin \theta
\simeq 0.94\sim 1,}
which shows that, on average, the multipolar anti-symmetric mode will scatter light very similarly to a point-like, dipolar atom.

During the RG flow, one can also encounter pairs of effective atoms that have a detuning of $\delta\omega_{12}=(\omega_1-\omega_2)/2$ with respect to each other. In order for these pairs to strongly interact and be renormalized, the pairwise interaction parameter should satisfy $\mathcal K_{12}>1$ , which is roughly equivalent to $|\delta\omega_{12}/G^{\text{near}}_{12}|\ll  1$. In this limit, one can readily extend the previous calculation to the case of two different atoms. In particular, after averaging the resonant cross-section of the (nearly) anti-symmetric, multipolar mode over all possible orientations, one finds
\eq{
\left \langle\dfrac{ \sigma_{\text{sc}}^-(\zeta)}{\sigma_{\text{sc}}}\right\rangle
=
\dfrac{5}{2}\left[1-(3\zeta^2+1)\dfrac{\arctanh \left(\frac{1}{\sqrt{5\zeta^2+2}}\right)}{\sqrt{5\zeta^2+2}}\right],
}
where $\zeta \equiv\delta\omega_{12}/(\rho_{12}G^{\text{near}}_{12})$, and which satisfies
\eq{
0.94 \lesssim \left \langle\dfrac{ \sigma_{\text{sc}}^-(\zeta)}{\sigma_{\text{sc}}}\right\rangle \leq 1.
}
Thus, we see that the multipolar mode of a pair of inequivalent atoms can also be well-approximated in its optical response by a single, electric dipolar atom.

}

\hypersetup{colorlinks=true,
urlcolor  = MidnightBlue
}

\addcontentsline{toc}{section}{References}
\bibliographystyle{my_ieeetr}
\bibliography{biblio}

\begin{thebibliography}{10}

\bibitem{Rustgi1961OpticalEv}
O.~P. Rustgi, J.~S. Nodvik, and G.~L. Weissler,
  \href{http://dx.doi.org/10.1103/PhysRev.122.1131}{``{Optical constants of
  germanium in the region 0-27 ev}'',} {\em Physical Review}, vol.~122,
  pp.~1131--1134, 5 1961.

\bibitem{Philipp1963OpticalSemiconductors}
H.~R. Philipp and H.~Ehrenreich,
  \href{http://dx.doi.org/10.1103/PhysRev.129.1550}{``{Optical properties of
  semiconductors}'',} {\em Physical Review}, vol.~129, pp.~1550--1560, 2 1963.

\bibitem{Walker1964UltravioletDiamond}
W.~C. Walker and J.~Osantowski,
  \href{http://dx.doi.org/10.1103/PhysRev.134.A153}{``{Ultraviolet optical
  properties of diamond}'',} {\em Physical Review}, vol.~134, p.~A153, 4 1964.

\bibitem{Lamy1977OpticalUltraviolet}
P.~L. Lamy, \href{http://dx.doi.org/10.1364/ao.16.002212}{``{Optical constants
  of crystalline and fused quartz in the far ultraviolet}'',} {\em Applied
  Optics}, vol.~16, p.~2212, 8 1977.

\bibitem{Aspnes1983DielectricEV}
D.~E. Aspnes and A.~A. Studna,
  \href{http://dx.doi.org/10.1103/PhysRevB.27.985}{``{Dielectric functions and
  optical parameters of Si, Ge, GaP, GaAs, GaSb, InP, InAs, and InSb from 1.5
  to 6.0 eV}'',} {\em Physical Review B}, vol.~27, pp.~985--1009, 1 1983.

\bibitem{Warren1984OpticalMicrowave}
S.~G. Warren, \href{http://dx.doi.org/10.1364/ao.23.001206}{``{Optical
  constants of ice from the ultraviolet to the microwave}'',} {\em Applied
  Optics}, vol.~23, p.~1206, 4 1984.

\bibitem{Papadopoulos1991OpticalDiamond}
A.~D. Papadopoulos and E.~Anastassakis,
  \href{http://dx.doi.org/10.1103/PhysRevB.43.5090}{``{Optical properties of
  diamond}'',} {\em Physical Review B}, vol.~43, pp.~5090--5097, 2 1991.

\bibitem{Kitamura2007OpticalTemperature}
R.~Kitamura, L.~Pilon, and M.~Jonasz,
  \href{http://dx.doi.org/10.1364/AO.46.008118}{``{Optical constants of silica
  glass from extreme ultraviolet to far infrared at near room temperature}'',}
  {\em Applied Optics}, vol.~46, pp.~8118--8133, 11 2007.

\bibitem{Jackson1998ClassicalElectrodynamics}
J.~D. Jackson,
  \href{https://www.wiley.com/en-us/Classical+Electrodynamics%2C+3rd+Edition-p-9780471309321}{{\em
  {Classical Electrodynamics}}}.
\newblock Wiley, 3rd~ed., 1998.

\bibitem{Grynberg2010IntroductionOptics}
G.~Grynberg {\em et~al.},
  \href{http://dx.doi.org/10.1017/cbo9780511778261}{{\em {Introduction to
  Quantum Optics}}}.
\newblock Cambridge University Press, 2010.

\bibitem{Fleischhauer1999RadiativeFormula}
M.~Fleischhauer and S.~F. Yelin,
  \href{http://dx.doi.org/10.1103/PhysRevA.59.2427}{``{Radiative atom-atom
  interactions in optically dense media: Quantum corrections to the
  Lorentz-Lorenz formula}'',} {\em Physical Review A}, vol.~59, pp.~2427--2441,
  3 1999.

\bibitem{Javanainen2016LightOptics}
J.~Javanainen and J.~Ruostekoski,
  \href{http://dx.doi.org/10.1364/oe.24.000993}{``{Light propagation beyond the
  mean-field theory of standard optics}'',} {\em Optics Express}, vol.~24,
  p.~993, 1 2016.

\bibitem{Manassah2012CooperativeShift}
J.~T. Manassah, \href{http://dx.doi.org/10.1364/aop.4.000108}{``{Cooperative
  radiation from atoms in different geometries: decay rate and frequency
  shift}'',} {\em Advances in Optics and Photonics}, vol.~4, p.~108, 6 2012.

\bibitem{Keaveney2012CooperativeThickness}
J.~Keaveney {\em et~al.},
  \href{http://dx.doi.org/10.1103/PhysRevLett.108.173601}{``{Cooperative Lamb
  Shift in an Atomic Vapor Layer of Nanometer Thickness}'',} {\em Physical
  Review Letters}, vol.~108, p.~173601, 4 2012.

\bibitem{Javanainen2014ShiftsSample}
J.~Javanainen {\em et~al.},
  \href{http://dx.doi.org/10.1103/PhysRevLett.112.113603}{``{Shifts of a
  resonance line in a dense atomic sample}'',} {\em Physical Review Letters},
  vol.~112, 3 2014.

\bibitem{Bromley2016CollectiveMedium}
S.~L. Bromley {\em et~al.},
  \href{http://dx.doi.org/10.1038/ncomms11039}{``{Collective atomic scattering
  and motional effects in a dense coherent medium}'',} {\em Nature
  Communications}, vol.~7, pp.~1--7, 3 2016.

\bibitem{Jenkins2016OpticalGases}
S.~D. Jenkins {\em et~al.},
  \href{http://dx.doi.org/10.1103/PhysRevLett.116.183601}{``{Optical Resonance
  Shifts in the Fluorescence of Thermal and Cold Atomic Gases}'',} {\em
  Physical Review Letters}, vol.~116, 5 2016.

\bibitem{Dobbertin2020CollectiveNanocavities}
H.~Dobbertin, R.~L{\"{o}}w, and S.~Scheel,
  \href{http://arxiv.org/abs/2003.06580}{``{Collective dipole-dipole
  interactions in planar nanocavities}'',} {\em arXiv:2003.06580}, 3 2020.

\bibitem{Schilder2016PolaritonicAtoms}
N.~J. Schilder {\em et~al.},
  \href{http://dx.doi.org/10.1103/PhysRevA.93.063835}{``{Polaritonic modes in a
  dense cloud of cold atoms}'',} {\em Physical Review A}, vol.~93, p.~063835, 6
  2016.

\bibitem{Schilder2017HomogenizationScatterers}
N.~J. Schilder {\em et~al.},
  \href{http://dx.doi.org/10.1103/PhysRevA.96.013825}{``{Homogenization of an
  ensemble of interacting resonant scatterers}'',} {\em Physical Review A},
  vol.~96, p.~013825, 7 2017.

\bibitem{Schilder2020Near-ResonantAtoms}
N.~Schilder {\em et~al.},
  \href{http://dx.doi.org/10.1103/PhysRevLett.124.073403}{``{Near-Resonant
  Light Scattering by a Subwavelength Ensemble of Identical Atoms}'',} {\em
  Physical Review Letters}, vol.~124, p.~073403, 2 2020.

\bibitem{Dicke1954CoherenceProcesses}
R.~H. Dicke, \href{http://dx.doi.org/10.1103/PhysRev.93.99}{``{Coherence in
  Spontaneous Radiation Processes}'',} {\em Physical Review}, vol.~93,
  pp.~99--110, 1 1954.

\bibitem{Gross1982Superradiance:Emission}
M.~Gross and S.~Haroche,
  \href{http://dx.doi.org/10.1016/0370-1573(82)90102-8}{``{Superradiance: An
  essay on the theory of collective spontaneous emission}'',} {\em Physics
  Reports}, vol.~93, pp.~301--396, 12 1982.

\bibitem{Roof2016ObservationAtoms}
S.~J. Roof {\em et~al.},
  \href{http://dx.doi.org/10.1103/PhysRevLett.117.073003}{``{Observation of
  Single-Photon Superradiance and the Cooperative Lamb Shift in an Extended
  Sample of Cold Atoms}'',} {\em Physical Review Letters}, vol.~117, p.~073003,
  8 2016.

\bibitem{Araujo2016SuperradianceRegime}
M.~O. Ara{\'{u}}jo {\em et~al.},
  \href{http://dx.doi.org/10.1103/PhysRevLett.117.073002}{``{Superradiance in a
  Large and Dilute Cloud of Cold Atoms in the Linear-Optics Regime}'',} {\em
  Physical Review Letters}, vol.~117, p.~073002, 8 2016.

\bibitem{Asenjo-Garcia2017ExponentialArrays}
A.~Asenjo-Garcia {\em et~al.},
  \href{http://dx.doi.org/10.1103/PhysRevX.7.031024}{``{Exponential Improvement
  in Photon Storage Fidelities Using Subradiance and "Selective Radiance" in
  Atomic Arrays}'',} {\em Phys. Rev. X}, vol.~7, p.~31024, 8 2017.

\bibitem{He2019GeometricEmission}
Y.~He {\em et~al.}, \href{http://arxiv.org/abs/1910.02289}{``{Geometric control
  of collective spontaneous emission}'',} {\em arXiv:1910.02289}, 10 2019.

\bibitem{Shahmoon2017CooperativeArrays}
E.~Shahmoon {\em et~al.},
  \href{http://dx.doi.org/10.1103/PhysRevLett.118.113601}{``{Cooperative
  Resonances in Light Scattering from Two-Dimensional Atomic Arrays}'',} {\em
  Physical Review Letters}, vol.~118, p.~113601, 3 2017.

\bibitem{Bettles2016EnhancedArray}
R.~J. Bettles, S.~A. Gardiner, and C.~S. Adams,
  \href{http://dx.doi.org/10.1103/PhysRevLett.116.103602}{``{Enhanced Optical
  Cross Section via Collective Coupling of Atomic Dipoles in a 2D Array}'',}
  {\em Physical Review Letters}, vol.~116, p.~103602, 3 2016.

\bibitem{Rui2020ALayerb}
J.~Rui {\em et~al.}, \href{http://dx.doi.org/10.1038/s41586-020-2463-x}{``{A
  subradiant optical mirror formed by a single structured atomic layer}'',}
  {\em Nature}, vol.~583, pp.~369--374, 7 2020.

\bibitem{Skipetrov2014AbsenceScatterers}
S.~Skipetrov and I.~Sokolov,
  \href{http://dx.doi.org/10.1103/PhysRevLett.112.023905}{``{Absence of
  Anderson Localization of Light in a Random Ensemble of Point Scatterers}'',}
  {\em Physical Review Letters}, vol.~112, p.~023905, 1 2014.

\bibitem{Skipetrov2019SearchField}
S.~E. Skipetrov and I.~M. Sokolov,
  \href{http://dx.doi.org/10.1103/PhysRevB.99.134201}{``{Search for Anderson
  localization of light by cold atoms in a static electric field}'',} {\em
  Physical Review B}, vol.~99, 4 2019.

\bibitem{Pellegrino2014ObservationEnsemble}
J.~Pellegrino {\em et~al.},
  \href{http://dx.doi.org/10.1103/PhysRevLett.113.133602}{``{Observation of
  Suppression of Light Scattering Induced by Dipole-Dipole Interactions in a
  Cold-Atom Ensemble}'',} {\em Physical Review Letters}, vol.~113, p.~133602, 9
  2014.

\bibitem{Jennewein2018CoherentTheory}
S.~Jennewein {\em et~al.},
  \href{http://dx.doi.org/10.1103/PhysRevA.97.053816}{``{Coherent scattering of
  near-resonant light by a dense, microscopic cloud of cold two-level atoms:
  Experiment versus theory}'',} {\em Physical Review A}, vol.~97, 5 2018.

\bibitem{Guerin2017LightEffects}
W.~Guerin, M.~T. Rouabah, and R.~Kaiser,
  \href{http://dx.doi.org/10.1080/09500340.2016.1215564}{``{Light interacting
  with atomic ensembles: collective, cooperative and mesoscopic effects}'',}
  {\em Journal of Modern Optics}, vol.~64, pp.~895--907, 5 2017.

\bibitem{Chomaz2012AbsorptionAnalysis}
L.~Chomaz {\em et~al.},
  \href{http://dx.doi.org/10.1088/1367-2630/14/5/055001}{``{Absorption imaging
  of a quasi-two-dimensional gas: a multiple scattering analysis}'',} {\em New
  Journal of Physics}, vol.~14, p.~055001, 5 2012.

\bibitem{Zhu2016LightMedia}
B.~Zhu {\em et~al.},
  \href{http://dx.doi.org/10.1103/PhysRevA.94.023612}{``{Light scattering from
  dense cold atomic media}'',} {\em Physical Review A}, vol.~94, p.~023612, 8
  2016.

\bibitem{Jenkins2016CollectiveExperiment}
S.~D. Jenkins {\em et~al.},
  \href{http://dx.doi.org/10.1103/PhysRevA.94.023842}{``{Collective resonance
  fluorescence in small and dense atom clouds: Comparison between theory and
  experiment}'',} {\em Physical Review A}, vol.~94, p.~023842, 8 2016.

\bibitem{Jennewein2016PropagationAtoms}
S.~Jennewein {\em et~al.},
  \href{http://dx.doi.org/10.1103/PhysRevA.94.053828}{``{Propagation of light
  through small clouds of cold interacting atoms}'',} {\em Physical Review A},
  vol.~94, no.~5, 2016.

\bibitem{Corman2017TransmissionAtoms}
L.~Corman {\em et~al.},
  \href{http://dx.doi.org/10.1103/PhysRevA.96.053629}{``{Transmission of
  near-resonant light through a dense slab of cold atoms}'',} {\em Phys. Rev.
  A}, vol.~96, p.~53629, 11 2017.

\bibitem{Jennewein2016CoherentCloud}
S.~Jennewein {\em et~al.},
  \href{http://dx.doi.org/10.1103/PhysRevLett.116.233601}{``{Coherent
  Scattering of Near-Resonant Light by a Dense Microscopic Cold Atomic
  Cloud}'',} {\em Physical Review Letters}, vol.~116, p.~233601, 6 2016.

\bibitem{Keaveney2012MaximalNanolayer}
J.~Keaveney {\em et~al.},
  \href{http://dx.doi.org/10.1103/PhysRevLett.109.233001}{``{Maximal Refraction
  and Superluminal Propagation in a Gaseous Nanolayer}'',} {\em Physical Review
  Letters}, vol.~109, p.~233001, 12 2012.

\bibitem{Levitov1990DelocalizationInteraction}
L.~S. Levitov,
  \href{http://dx.doi.org/10.1103/PhysRevLett.64.547}{``{Delocalization of
  vibrational modes caused by electric dipole interaction}'',} {\em Physical
  Review Letters}, vol.~64, p.~547, 1 1990.

\bibitem{Fisher1994RandomChains}
D.~S. Fisher, \href{http://dx.doi.org/10.1103/PhysRevB.50.3799}{``{Random
  antiferromagnetic quantum spin chains}'',} {\em Physical Review B}, vol.~50,
  pp.~3799--3821, 8 1994.

\bibitem{Damle2000DynamicsChains}
K.~Damle, O.~Motrunich, and D.~A. Huse,
  \href{http://dx.doi.org/10.1103/PhysRevLett.84.3434}{``{Dynamics and
  transport in random antiferromagnetic spin chains}'',} {\em Physical Review
  Letters}, vol.~84, pp.~3434--3437, 4 2000.

\bibitem{Motrunich2000Infinite-randomnessPoints}
O.~Motrunich {\em et~al.},
  \href{http://dx.doi.org/10.1103/PhysRevB.61.1160}{``{Infinite-randomness
  quantum Ising critical fixed points}'',} {\em Physical Review B}, vol.~61,
  pp.~1160--1172, 1 2000.

\bibitem{Refael2004EntanglementDimension}
G.~Refael and J.~E. Moore,
  \href{http://dx.doi.org/10.1103/PhysRevLett.93.260602}{``{Entanglement
  entropy of random quantum critical points in one dimension}'',} {\em Physical
  Review Letters}, vol.~93, p.~260602, 12 2004.

\bibitem{Igloi2005StrongSystems}
F.~Igl{\'{o}}i and C.~Monthus,
  \href{http://dx.doi.org/10.1016/j.physrep.2005.02.006}{``{Strong disorder RG
  approach of random systems}'',} {\em Physics Reports}, vol.~412,
  pp.~277--431, 6 2005.

\bibitem{Vosk2013Many-bodyPoint}
R.~Vosk and E.~Altman,
  \href{http://dx.doi.org/10.1103/PhysRevLett.110.067204}{``{Many-body
  localization in one dimension as a dynamical renormalization group fixed
  point}'',} {\em Physical Review Letters}, vol.~110, p.~067204, 2 2013.

\bibitem{Refael2013StrongTransition}
G.~Refael and E.~Altman,
  \href{http://dx.doi.org/10.1016/j.crhy.2013.09.005}{``{Strong disorder
  renormalization group primer and the superfluid-insulator transition}'',}
  {\em Comptes Rendus Physique}, vol.~14, pp.~725--739, 10 2013.

\bibitem{Lagendijk1996ResonantLight}
A.~Lagendijk and B.~A. Van~Tiggelen,
  \href{http://dx.doi.org/10.1016/0370-1573(95)00065-8}{``{Resonant multiple
  scattering of light}'',} {\em Physics Report}, vol.~270, pp.~143--215, 5
  1996.

\bibitem{Fayard2015IntensityPatterns}
N.~Fayard {\em et~al.},
  \href{http://dx.doi.org/10.1103/PhysRevA.92.033827}{``{Intensity correlations
  between reflected and transmitted speckle patterns}'',} {\em Physical Review
  A}, vol.~92, p.~033827, 9 2015.

\bibitem{Cottier2019MicroscopicLight}
F.~Cottier {\em et~al.},
  \href{http://dx.doi.org/10.1103/PhysRevLett.123.083401}{``{Microscopic and
  Macroscopic Signatures of 3D Anderson Localization of Light}'',} {\em
  Physical Review Letters}, vol.~123, 8 2019.

\bibitem{Binninger2019NonlinearCloud}
T.~Binninger {\em et~al.},
  \href{http://dx.doi.org/10.1103/PhysRevA.100.033816}{``{Nonlinear quantum
  transport of light in a cold atomic cloud}'',} {\em Physical Review A},
  vol.~100, 9 2019.

\bibitem{Novotny2009PrinciplesNano-optics}
L.~Novotny and B.~Hecht, \href{http://dx.doi.org/10.1017/CBO9780511794193}{{\em
  {Principles of nano-optics}}}.
\newblock Cambridge University Press, 1 2009.

\bibitem{GarciaDeAbajo2007CollectiveMatter}
F.~J. Garc{\'{i}}a De~Abajo,
  \href{http://dx.doi.org/10.1364/oe.15.011082}{``{Collective oscillations in
  optical matter}'',} {\em Optics Express}, vol.~15, p.~11082, 9 2007.

\bibitem{DeVries1998PointWaves}
P.~De~Vries, D.~V. Van~Coevorden, and A.~Lagendijk,
  \href{http://dx.doi.org/10.1103/revmodphys.70.447}{``{Point scatterers for
  classical waves}'',} {\em Reviews of Modern Physics}, vol.~70, pp.~447--466,
  4 1998.

\bibitem{Shapiro1986LargeMedia}
B.~Shapiro, \href{http://dx.doi.org/10.1103/PhysRevLett.57.2168}{``{Large
  intensity fluctuations for wave propagation in random media}'',} {\em
  Physical Review Letters}, vol.~57, pp.~2168--2171, 10 1986.

\bibitem{Manzoni2018OptimizationArrays}
M.~T. Manzoni {\em et~al.},
  \href{http://dx.doi.org/10.1088/1367-2630/aadb74}{``{Optimization of photon
  storage fidelity in ordered atomic arrays}'',} {\em New J. Phys.}, vol.~20,
  p.~83048, 8 2018.

\bibitem{Chang2012CavityMirrors}
D.~E. Chang {\em et~al.},
  \href{http://dx.doi.org/10.1088/1367-2630/14/6/063003}{``{Cavity QED with
  atomic mirrors}'',} {\em New Journal of Physics}, vol.~14, p.~63003, 6 2012.

\bibitem{Davis1995Bose-EinsteinAtoms}
K.~B. Davis {\em et~al.},
  \href{http://dx.doi.org/10.1103/PhysRevLett.75.3969}{``{Bose-Einstein
  condensation in a gas of sodium atoms}'',} {\em Physical Review Letters},
  vol.~75, pp.~3969--3973, 11 1995.

\bibitem{Anderson1995ObservationVapor}
M.~H. Anderson {\em et~al.},
  \href{http://dx.doi.org/10.1126/science.269.5221.198}{``{Observation of
  Bose-Einstein condensation in a dilute atomic vapor}'',} {\em Science},
  vol.~269, pp.~198--201, 7 1995.

\bibitem{Born1999PrinciplesOptics}
M.~Born and E.~Wolf, \href{http://dx.doi.org/10.1017/CBO9781139644181}{{\em
  {Principles of Optics}}}.
\newblock Cambridge University Press, 7~ed., 1999.

\bibitem{Allard1982TheLines}
N.~Allard and J.~Kielkopf,
  \href{http://dx.doi.org/10.1103/RevModPhys.54.1103}{``{The effect of neutral
  nonresonant collisions on atomic spectral lines}'',} {\em Reviews of Modern
  Physics}, vol.~54, pp.~1103--1182, 10 1982.

\bibitem{Bellando2014CooperativeHamiltonian}
L.~Bellando {\em et~al.},
  \href{http://dx.doi.org/10.1103/PhysRevA.90.063822}{``{Cooperative effects
  and disorder: A scaling analysis of the spectrum of the effective atomic
  Hamiltonian}'',} {\em Physical Review A}, vol.~90, p.~063822, 12 2014.

\bibitem{Skipetrov2016RedLocalization}
S.~E. Skipetrov and J.~H. Page,
  \href{http://dx.doi.org/10.1088/1367-2630/18/2/021001}{``{Red light for
  Anderson localization}'',} {\em New Journal of Physics}, vol.~18, 1 2016.

\bibitem{Skipetrov2018Ioffe-RegelScatterers}
S.~E. Skipetrov and I.~M. Sokolov,
  \href{http://dx.doi.org/10.1103/PhysRevB.98.064207}{``{Ioffe-Regel criterion
  for Anderson localization in the model of resonant point scatterers}'',} {\em
  Physical Review B}, vol.~98, p.~064207, 8 2018.

\bibitem{Gorshkov2007UniversalMedia}
A.~V. Gorshkov {\em et~al.},
  \href{http://dx.doi.org/10.1103/PhysRevLett.98.123601}{``{Universal Approach
  to Optimal Photon Storage in Atomic Media}'',} {\em Physical Review Letters},
  vol.~98, p.~123601, 3 2007.

\bibitem{Hammerer2010QuantumEnsembles}
K.~Hammerer, A.~S. S{\o}rensen, and E.~S. Polzik,
  \href{http://dx.doi.org/10.1103/RevModPhys.82.1041}{``{Quantum interface
  between light and atomic ensembles}'',} {\em Rev. Mod. Phys.}, vol.~82,
  pp.~1041--1093, 4 2010.

\bibitem{Gorshkov2011Photon-PhotonBlockade}
A.~V. Gorshkov {\em et~al.},
  \href{https://link.aps.org/doi/10.1103/PhysRevLett.107.133602}{``{Photon-Photon
  Interactions via Rydberg Blockade}'',} {\em Phys. Rev. Lett.}, vol.~107,
  p.~133602, 9 2011.

\bibitem{Naccarato2019SearchingStudy}
F.~Naccarato {\em et~al.},
  \href{http://dx.doi.org/10.1103/PhysRevMaterials.3.044602}{``{Searching for
  materials with high refractive index and wide band gap: A first-principles
  high-throughput study}'',} {\em Physical Review Materials}, vol.~3, 4 2019.

\bibitem{Shubnic2020HighNanophotonics}
A.~A. Shubnic {\em et~al.},
  \href{http://dx.doi.org/10.1515/nanoph-2020-0416}{``{High refractive index
  and extreme biaxial optical anisotropy of rhenium diselenide for applications
  in all-dielectric nanophotonics}'',} {\em Nanophotonics}, vol.~9,
  pp.~4737--4742, 11 2020.

\bibitem{Newton1976OpticalBeyond}
R.~G. Newton, \href{http://dx.doi.org/10.1119/1.10324}{``{Optical theorem and
  beyond}'',} {\em American Journal of Physics}, vol.~44, pp.~639--642, 7 1976.

\bibitem{Draine1988TheGrains}
B.~T. Draine, \href{http://dx.doi.org/10.1086/166795}{``{The discrete-dipole
  approximation and its application to interstellar graphite grains}'',} {\em
  The Astrophysical Journal}, vol.~333, p.~848, 10 1988.

\bibitem{Sosa2003OpticalShapes}
I.~O. Sosa, C.~Noguez, and R.~G. Barrera,
  \href{http://dx.doi.org/10.1021/jp0274076}{``{Optical properties of metal
  nanoparticles with arbitrary shapes}'',} {\em Journal of Physical Chemistry
  B}, vol.~107, pp.~6269--6275, 7 2003.

\bibitem{Evlyukhin2011MultipoleApproximation}
A.~B. Evlyukhin, C.~Reinhardt, and B.~N. Chichkov,
  \href{http://dx.doi.org/10.1103/PhysRevB.84.235429}{``{Multipole light
  scattering by nonspherical nanoparticles in the discrete dipole
  approximation}'',} {\em Physical Review B}, vol.~84, p.~235429, 12 2011.

\bibitem{Evlyukhin2013MultipoleSurface}
A.~B. Evlyukhin {\em et~al.},
  \href{http://dx.doi.org/10.1364/josab.30.002589}{``{Multipole analysis of
  light scattering by arbitrary-shaped nanoparticles on a plane surface}'',}
  {\em Journal of the Optical Society of America B}, vol.~30, p.~2589, 10 2013.

\bibitem{Alaee2020KerkerAntennasb}
R.~Alaee {\em et~al.},
  \href{http://dx.doi.org/10.1103/physrevresearch.2.043409}{``{Kerker effect,
  superscattering, and scattering dark state in atomic antennas}'',} {\em Phys.
  Rev. Research}, vol.~2, p.~043409, 8 2020.

\end{thebibliography}

\end{document}